\documentclass[twocolumn,english,aps,superscriptaddress,pra]{revtex4-1}

\usepackage{xcolor}
\usepackage{array}
\usepackage[latin9]{inputenc}
\usepackage{amsmath}
\usepackage{amssymb}
\usepackage{graphicx}
\usepackage{babel}
\usepackage{mathrsfs}
\usepackage{amsfonts}
\usepackage{epstopdf}
\usepackage{float}
\usepackage{multirow}
\usepackage{color}
\usepackage{natbib}

\begin{document}
\title{Emergent $U(1)$ Symmetries in Gapless Fermionic Superfluids or Superconductors}

\author{Fei Zhou}

\affiliation{Department of Physics and Astronomy, University of British Columbia, 6224 Agricultural Road, Vancouver, BC, V6T 1Z1, Canada}

\begin{abstract}
A superfluid spontaneously breaks the usual $U(1)$ symmetry because of condensation. 
In this article, we illustrate six linearly independent families emergent $U(1)$ symmetries that naturally appear in infrared limits in a broad class of generic gapless topological superfluids
(that either belong to a stable phase or are quantum critical).
In gapless states we have considered, emergent $U(1)$ symmetry groups are embedded in an $Spin(4)=SU(2) \otimes SU(2)$ group that double covers (and algebraically is isomorphic to) an $SO(4)$ group.
All $U(1)$ charges associated with symmetries are further invariant under an $SU(2)$ spin group or an equivalent of it but always break pre-existing higher space-time Lorentz symmetry of $SO(3,1)$ group. 
Emergent $U(1)$ symmetries can be further spontaneously broken only if interactions are strong enough and resultant strong coupling states become fully gapped.
However if states remain gapless, emergent $U(1)$ symmetries are always present, despite that these states may exhibit much lower space-time symmetries
compared to their weakly interacting gapless Lorentz symmetric counter parts. In the limit of our interests, we have identified all possible gapless real fermions with or without Lorentz symmetries  and find that they all display emergent $U(1)$ symmetries in the infrared limit.  We argue emergent $U(1)$ symmetries are intrinsic in a broad class of interacting gapless superfluid or superconducting states and are typically well defined in high dimensions where there are infrared stable fixed points dictating emergent properties.
\end{abstract}

\date{\today}
\maketitle

 \section{Introduction}
  
 It is well known that the $U(1)$ symmetry is always spontaneously broken in superfluids (or to a large extent in superconductors). So in either superfluids or superconductors, charges are not conserved because of condensates in ground states and low energy dynamics are generically characterized by emergent real fermions rather than conventional complex fermions. These emergent particles are not only crucial in discussions of topological
 states, but also play critical roles in studies of topological quantum criticality as well as  in applications to quantum technologies\cite{Fu08,Schnyder08,Kitaev09,Qi10,Volovik88,Volovik03,Read00,Qi11,Bernevig13,Hasan10}.
 
Naturally related to emergent fermions are the emergent symmetries of these fields or particles. In a broad class of topological states where not only $U(1)$ symmetry but also other continuous symmetries such as rotation one are broken, there can be surprising emergent symmetries of very large groups in the low energy subspace. These symmetries can be totally unexpected as one usually does not anticipate their appearance in low energy scales 
solely based on microscopic considerations. Their existence at first sight even appears to be inconsistent with underlying symmetries. Nevertheless, they do emerge in many physical systems.

Consider a topologically non-trivial 3D $p$-wave superfluid or superconductor with time reversal symmetry. A well-known concrete example for this can be a Balian-Werthamer (BW) state of $p$-wave superfluid or Helium-3 B phase\cite{Leggett75} that breaks $U(1)$ symmetry
and spatial rotation symmetry (but with spin-spatial rotation symmetry intact). In strong coupling limits where pairing is very strong or the one-particle band is flat so that Fermi surface effects are completely suppressed, the low energy dynamics can be fully described by real fermions with an emergent Lorentz symmetry while underlying fermions are entirely non-relativistic\cite{Volovik03,Qi11}. In this case, strong interactions lead to highly surprising dynamics with very high symmetry.

These emergent symmetries have played very important roles in previous studies of topological quantum criticality in superfluids. In fact, they are part of symmetry groups that have been used to identify universality classes, apart from
discrete global symmetries\cite{Yang19,Yang21,Zhou22}.  While continuous emergent symmetries usually lead to concrete scaling symmetries, discrete global symmetries define the number of relevant low energy degrees in fermion fields or central charges. Both are crucial in studies of thermodynamics and dynamics near topological critical points.

The emergent symmetries in gapless limits are also usually further higher than adjacent gapped phases. This can be understood in terms of mass operators in gapped phases.
Presence of these additional mass operators always lowers pre-existing symmetries of their gapless counterparts . From this point of view, gapless states, either critical or belonging to stable phases usually
have the highest emergent symmetries when compared with surrounding gapped phases. This article is mainly focused on those highly symmetric gapless states.
Topological and dynamic stability of various gapless states had been focuses in many previous studies\cite{Sato17,Wen02,Sato06,Beri10,Kobayashi14,Zhao16,Schnyder11,Wan11,Burkov11,Burkov18,Armitage18,Meng12,Cho12,Grover14}
and we will refer them to those original research.

Below we will mainly emphasize two important aspects of translationally invariant gapless superfluids or superconductors.
One is about connections between physically different gapless states that exhibit the Lorentz invariant dynamics with  the same partition functions, and relations between emergent $U(1)$ symmetries in different systems.
This is effectively to classify and establish equivalent families of apparently different gapless states with different emergent $U(1)$ symmetries unique to gapless states.   

More specially, each unique emergent symmetry in a specific gapless state is associated with invariance under a particular transformation induced by a generator. 
In the limit of our interests, all these generators turn out to be purely imaginary or anti-symmetric Hermitian operators.
The complete set of such operators can be classified into ones of an $SU(2)$ spinor rotation subgroup of Lorentz transformation generated by
$S^{ij}$, $i\neq j=x,y,z$, and ones of its dual $SU(2)$ group induced by $S_D^{ij}$, $i\neq j=x,y,z$ that are mutually commuting with $S^{ij}$. 
The combination of these two $SU(2)$ groups, $SU(2)\otimes SU(2)$, forms an $Spin(4)$ group of {\em real fermions} that  double covers an $SO(4)$ group  with isomorphic algebras.

Emergent $U(1)$ symmetry groups are thus embedded in this $Spin(4)=SU(2) \otimes SU(2)$ group.
The conserved charge associated with the emergent symmetry is always rotationally invariant under the action of $S^{ij}$ and has to be represented by one
of the generators in the dual group, $S_D^{ij}$.
However, if gapless states become gapped, we find the unique emergent $U(1)$ symmetry associated with gapless liquids is always broken although the Lorentz symmetry always remains.

The second objective is to explore relations between gapless states with higher symmetries including $SO(3,1)$ Lorentz symmetry and more general gapless states with much lower symmetries i.e. without the Lorentz symmetry. 
In the limit we have considered, the emergent $U(1)$ symmetry associated to the dual group generated by $S^{ij}_D$, $i,j=x,y,z$ and its conserved charges are always present 
as far as states are gapless, disregarding the Lorentz symmetry. Quantum dynamics of gapless liquids strongly depend on whether they 
display a full $SO(3,1)$ Lorentz symmetry, or there is only a lower $SO(3)$ or $SO(2)$ rotation symmetry.
However, it appears that they all share the same emergent $U(1)$ symmetries associated with the dual group of $S_D^{ij}$, disregarding differences in dynamics.   
We conjecture such an emergent symmetry to be fundamental to  many gapless states in superfluids or superconductors.

The article is organized as follows.
In section II, we introduce effective field theories for discussions of gapless states in superfluids or superconductors. We argue that Lorentz symmetry naturally emerges when only the most relevant terms are included. 
We also explicitly list the symmetry properties (complex vs real) of Dirac operators and Lorentz group generators in the real fermion representation used in this article.
In section III, we illustrate the structure of unitary rotations of the real fields.
These unitary rotations which are represented by {\em real} matrices only form an $Spin(4)$ group.  
In section IV, we briefly summarize the main results.

In section V, we examine all the generators which appear in the $Spin(4)$ group algebras and show that one can always identify one of them as a $U(1)$ symmetry generator in any limits.
We therefore illustrate that $U(1)$ symmetry as a robust feature in gapless superfluid states near a stable infrared fixed point and also discuss to what extend they can emerge in the infrared limit.
In section VI, we discuss the emergent $U(1)$-charges in a few concrete gapless superfluid states such as strong coupling $p$-wave superfluids and nodal point phases, and explore possible physical consequences
of emergent {\em charges}.
In section VII, we illustrate that the emergent infrared $U(1)$ symmetry can also be an asymptotic symmetry in more generic interacting gapless states.
Moving into higher energies, we however find that $U(1)$ can naturally evolve into $Z_2$ symmetries in gapless liquids.
In section VIII, we discuss what happens to the emergent $U(1)$ symmetries in gapless superfluids in more general cases without the Lorentz symmetry. 
We further show that by condensing one of generalized mass operators that break the Lorentz symmetry, the gapless liquids can be further transformed into other gapless states with rotational symmetries only, lower than the Lorentz symmetry. 
Nevertheless, the $U(1)$ symmetry is still intact in all gapless liquids we have examined. It can be broken only when the Lorentz symmetry remains unbroken but states become fully gapped. 
In section IX, we conclude our studies and point out a few open questions.
Some of detailed analyses are presented in Appendices so the discussions in the article are self-contained
In Appendix A, we show explicitly the structure of $Spin(4)$ group which double covers $SO(4)$ and lead the desired rotations of real fermions. We also defines a specific basis for the constructions of {\em EFT}s for different physical systems in this article. In Appendix B, we show various mappings between effective field theories for different physical systems and hence establish an equivalence between different $U(1)$-symmetries or $U(1)$ charges from the point of view of {\em EFT}s.

\section{Effective Field Theories (EFTs) for Relativistic Real Fermions}

\subsection{Effective Field Theories}
To understand interaction dynamics in superfluids and superconductors especially topological aspects, it is often very convenient to employ the real fermion representation to faithfully represent the intrinsic charge conjugation symmetry.
Below we are going to use the real fermion representation to explore simple relations between physically very different superconducting states or superfluids. The purpose is to show that dynamics in many different systems are 
entirely equivalent and are universal and therefore studying one is equivalent to exploring the whole equivalent class.

Without loosing generality, we can cast an interacting Hamiltonian of {\em gapless real fermions} in the following infrared form,

\begin{eqnarray}
H_{eff} &=& \frac{1}{2} \int d{\bf r} [ \chi^T({\bf r}){\bf  \alpha} \cdot {i\nabla} \chi ({\bf r})
\nonumber \\
&+& g_1 \chi^T \beta_1 \chi \chi^T  \beta_1 \chi +g_2 \chi^T  \beta_2 \chi  \chi^T \beta_2 \chi +... ]
\label{Eff}
\end{eqnarray}
where ${\bf \alpha}=(\alpha_x,\alpha_y,\alpha_z)$. And $\alpha_{i}$, $i=x,y,z$ and $\beta_{1,2}$ are mutually anti-commuting Hermitian matrices.
That is,

\begin{eqnarray}
&& \{ \alpha_i, \alpha_j \}=2\delta_{i,j}, \{\beta_m,\beta_n \}=2\delta_{m,n}, \{ \alpha_i,\beta_m \}=0; \nonumber \\
&& \alpha^\dagger_i=\alpha_i, \beta^\dagger_m=\beta_m, i=x,y,z; m=1,2.
\label{Algebra}
\end{eqnarray}

We will restrict to real fermions with four components, which turns out to be a minimum number of degrees of freedom for our discussions;

\begin{eqnarray}
&& \chi^T=(\chi_1,\chi_2,\chi_3,\chi_4), \chi^\dagger_i({\bf r}) =\chi_i ({\bf r}),  \{ \chi_i, \chi_j \}=\delta_{ij}.
\end{eqnarray}
We have also chosen to introduce two interactions, $g_{1,2}$ for later discussions on emergent symmetries, although for the four component real fermions, these two operators turn out to be always equivalent.  
So for the convenience of the rest of discussions in this section, we first set $g_1=g_0$ and $g_2=0$, and $\beta_1=\beta_0$; this is equivalent to a procedure of gauge-fixing in the case of emergent global $U(1)$ 
symmetries or dropping a less relevant channel in the case of emergent $Z_2$ symmetries. One can also verify that to be fully consistent with real fermion representations, $\beta_0$ also has to be purely imaginary and anti-symmetric;
otherwise the four fermion operator becomes nullified. And $\alpha$ matrices are real and symmetric ones. That is,

\begin{eqnarray}
&& \alpha_i^T=\alpha_i=\alpha_i^*, \beta^T_0=-\beta_0=-\beta_0^*; i=x,y,z.
\end{eqnarray}

In presenting Eq.(\ref{Eff}), we have only kept most relevant kinetic and interaction terms.

1) We have muted the terms bilinear in $\chi$ but higher order in $\nabla$,  i.e. $\chi \nabla^2\chi, \chi\nabla^3\chi$ (as $...$) as they are less relevant in the infrared limit. We have also re-adjusted the velocity of fermions along $x,y,z$ direction to be equal by a trivial rescaling of $x,y,z$ and keep the terms linear in $\nabla$. In very special cases when one of the velocities is exactly zero and the leading terms involve $\nabla^2$, the effective theory shall be of quantum Lifshitz majorana fields. The physics of those was discussed in Ref.\cite{Yang21} and will not be the focus here. However, most of the conclusions derived here shall also be applicable to those models in the infrared limit as generally, $\nabla^2$ terms are less relevant than terms linear in $\nabla$ we have kept here. Eq.(\ref{Eff}) is a generic infrared theory for a broad class of gapless superfluids or superconductors with intrinsic (relativistic) particle-hole symmetries and with dynamics captured by leading linear-in-$\nabla$ terms.  
 
 From the point of view of scaling dimensions, Lifshitz majorana fields are less generic as they require fine tuning of fermion velocity to zero so that more relevant kinetic terms vanish identically.
 Therefore, unless such an effective field theory violates additional symmetry constraints, Eq.(\ref{Eff}) shall be considered to be a more generic form of low energy interacting real fermions which is naturally Lorentz invariant.
 However, if physical systems are further constrained by other continuous symmetries such as $SO(2)$ or $SO(3)$ spatial rotational ones, effective fields then have to fall into those Lifshitz classes discussed before \cite{Yang21}.

2) We also have muted the four fermion terms involving additional  $\nabla$ (as $...$ in line 2) as they are also less relevant compared to the four-fermion terms kept.  

This generic form of gapless real fermions naturally have a very high space-time Lorentz symmetry. We will take this as a starting point of discussions on gapless fermions.
For the purpose of emergent $U(1)$ symmetries to be discussed later, it turns out that this naturally emergent high space-time symmetry is un-essential at all and the emergent $U(1)$ symmetry can appear in all other Lorentz non-invariant gapless fermion systems including the ones with Fermi surfaces. 
Nevertheless, this highly symmetric  limit is the most convenient focal point where other gapless fermions can be easily related to and for that reason, we will spend quite bit efforts to examine this limit first before extending to 
other less symmetric but equally interesting cases. 

For the same reason, we also first restrict ourselves to Lorentz symmetric interactions $g_{1,2}$ only and do not consider less relevant interactions that break the Lorentz symmetry.
Later, we will see algebras in Eq.(\ref{Algebra}) are sufficient for emergent Lorentz symmetry, even when interactions are strong.

\subsection{Emergent Lorentz Symmetry}

The  generic model for infrared physics defined by ${\bf \alpha}$, $\beta_0$  has an emergent Lorentz symmetry as in the standard relativistic theories.
For instance, one can then construct the $4\times 4$ gamma matrices and Lorentz group generators in a standard way.

\begin{eqnarray}
&& \Gamma_0=\beta_0, \Gamma_i =\beta_0 {\bf \alpha}_i, i=x,y,z. \nonumber \\ 
&& \{ \Gamma_\mu, \Gamma_\nu \}=g_{\mu\nu}, \mu, \nu=0,x,y,z. 
\end{eqnarray}

The Lorentz boost are generated by $S^{0i}$, $i=x,y,z$ and rotations are  $S^{ij}$, $i\neq j=x,y,z$. Together they generate the $SO(3,1)$ Lorentz group. They can be explicitly defined in terms of  $\alpha_i, \beta_0$, $i=x,y,z$;  

\begin{eqnarray}
&& S^{0i}=\frac{i}{2} \alpha_i, S^{ij}=-S^{ji}=-\frac{i}{4} [\alpha_i,\alpha_j];i, j=x,y,z.   
\label{S}
\end{eqnarray}
Furthermore, the boost operators are anti-hermitian, symmetric and purely imaginary; the rotation ones as usually are hermitian and antisymmetric and purely imaginary, i.e., 

\begin{eqnarray}
{S^{0i} }^\dagger & =&-S^{0i}=-{S^{0i}}^T={S^{0j}}^*;  \nonumber \\
 {S^{ij}}^\dagger &=& S^{ij}=-{S^{ij}}^T=-{S^{ij}}^*, i\neq j=x,y,z.
\end{eqnarray}

It is important to notice that the Lorentz $SO(3,1)$ group can be fully generated by $S^{0i}$ and $S^{ij}$ which are independent of the choices of $\beta$ as far as Eq.(\ref{Algebra}) are satisfied. 
So the Lorentz symmetry can emerge even when $g_2$ is nonzero and interactions of $\beta_2$-type are present. 
From now on, we can relax the condition of $g_2=0$ and again keep both $g_{1,2}$. 

We can further define $s_i=\frac{1}{2}\epsilon_{ijk}S^{jk}$. Following the general anti-commuting relations between $\alpha_i$ in Eq.(\ref{Algebra}) , we verify that $s_i$, $i=x,y,z$, being anti-symmetric and purely imaginary,  satisfy the simple $SU(2)$ algebra, i.e. $S^{ij}$ form the algebras of an $SU(2)$ spinor subgroup in the Lorentz group of $SO(3,1)$. Together with $t_i=S^{0i}$, $\{ s_i, t_i\}$, $i=x,y,z$ in Eq.(\ref{S}) indeed lead to the $SO(3,1)$ algebras of the Lorentz group, 

\begin{eqnarray}
[ s_i, s_j ] &=& i\epsilon_{ijk} s_k,  
[ t_i, t_j ]= -i\epsilon_{ijk} s_k,
[ s_i, t_j ] = i\epsilon_{ijk} t_k,  \nonumber \\
\end{eqnarray}
where $i,j.k=x,y,z$.

For an operator to be Lorentz invariant, the operator has to commute with the rotation subgroup generators $S^{ij}$, $i\neq j=x,y,z$ but anti-commute with the boost generators of $S^{0i}$.
Following Eq.(\ref{S}), any operator anti-commuting with $\alpha_i$ or the booster operators $S^{0i}$ is also rotationally invariant; so it is further Lorentz invariant. For this reason, $\beta_0$ is obviously Lorentz invariant by this definition,

\begin{eqnarray}
\{ \beta_0, S^{0i} \}=[ \beta_0, S^{ij} ]=0, i \neq j=x,y,z.
\end{eqnarray}

\begin{figure}
\includegraphics[width=8cm]{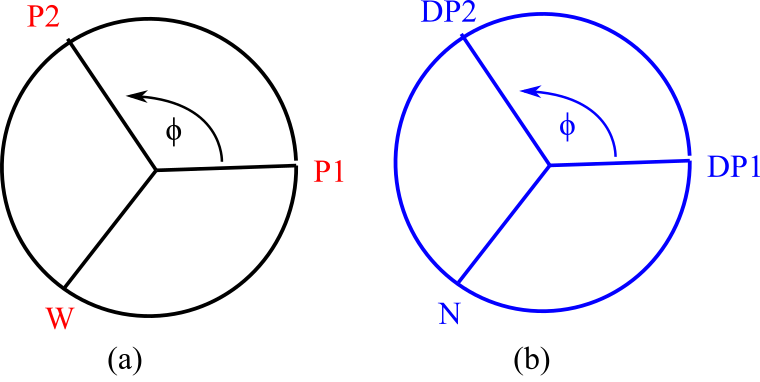}
\centering
\caption{ \footnotesize Equivalence between {\em real fermions} dynamics in gapless superfluids or superconductors. $\phi$ is the rotation angle of an $SU(2)$ rotation defined in Eq.(\ref{ROT}) where ${\bf n}=1/\sqrt{3}(1,1,1)$.
a). {\em P1} refers to a time reversal invariant $p$-wave state; {\em P2} as state {\em P1} but with an additional $\pi/2$-phase shift to further break the time reversal symmetry; {\em W} is related to a real fermion  representation of complex Weyl fermions. b). The $\tau-\sigma$ duals of the dynamics in a). {\em DP1(2)} is dual to {\em P1(2)} and the dual of state {\em W} in a) is {\em N}, that can be representing dynamics of a nodal point phase.  }
\label{EMS1}
\end{figure}

\section{$Spin(4)$ group in {\em EFTs} of real fermions}

To construct effective field theories ({\em EFT}) of {\em real fermions} and find a specific representation for $\alpha_i, \beta_0$, $i=x,y,z$, we shall group all hermitian matrices into two big categories:

1) real and symmetric ones that can only appear along with gradient operators or odd powers of gradient operators;

2) imaginary and antisymmetric ones that can only appear along with mass terms or even powers of gradient operators.

In this article, we are mainly interested in four component real fermion field theories, which is equivalent to one-half of standard three dimensional four-component Dirac fermions.
The $4\times 4$ hermitian matrices form a representation of  ${su}(4)$, a algebraic group (note $SU(4)$ group itself is not a symmetry group for real fermions discussed here) that contains three independent $SU(2)$ subgroup algebras; these subgroups play important roles in later discussions.
We construct these operators via a tensor product of two $SU(2)$ group algebras, one acting on the particle-anti-particle Nambu space by $\tau_{1,2,3}$ and the other acting on the standard spin space by $\sigma_{x,y,z}$. 
 The 15 hermitian matrices constructed in this way are
 
 \begin{eqnarray}
 && \mbox{Six Imaginary and Asymmetric Operators} \nonumber \\ 
 && K_1=\tau_y \otimes \sigma_x,  K_2=\tau_y \otimes \sigma_z, K_3=1 \otimes \sigma_y,  \nonumber \\
 && F_1=\tau_x \otimes \sigma_y, F_2=\tau_z \otimes \sigma_y, F_3=\tau_y \otimes 1; \nonumber \\
 \label{Asym11}
 && \mbox{Nine Real and Symmetric Operators} \nonumber \\
 && \tau_x \otimes \sigma_x, \tau_x \otimes \sigma_z, \tau_z \otimes 1,  \nonumber \\
 && \tau_z \otimes \sigma_x, \tau_z \otimes \sigma_z, \tau_x \otimes 1,  \nonumber \\
 &&1 \otimes \sigma_x, 1 \otimes \sigma_z,  \tau_y \otimes \sigma_y.
 \end{eqnarray}

 Note that six asymmetric operators form two independent $SU(2)$ subgroup algebras isomorphically and respectively;
 \begin{eqnarray}
&& [K_i, K_j]=i2\epsilon_{ijk} K_k, [F_i, F_j]=i2\epsilon_{ijk} F_k,\nonumber \\
 && [K_i, F_j]=0, i,j=1,2,3.
 \label{KF}
 \end{eqnarray}
 Unlike the standard Pauli matrices, these purely {\em imaginary} hermitian matrices form a {real} representation of $SU(2)$ rotations
and can be employed to perform $SU(2)$ rotations of {\em real} fermion fields. They play paramount roles in emergent symmetries and spontaneously symmetry breaking of them.
{\em One can show in our {\em EFT}, all the unitary transformations that leave real fermions in a real representation can be generated by these six operators, i.e. an $Spin(4)$ group algebra.} 
We will name the symmetries associated with real $SU(2)$ rotations generated by $K_i , i=1,2,3$ as the {\em $\sigma$-symmetry}, and the ones associated with $SU(2)$ rotations generated by $F_{i}, i=1,2,3$ as the
{\em  $\tau$-symmetry}.  As evident, these two $SU(2)$ groups are invariant under an interchange of $\sigma \rightarrow \tau$, $\tau \rightarrow \sigma$ leading to a {\em $\sigma-\tau$ duality} (see Appendix A for details).

Two $SU(2)$ groups also double cover  an $SO(4)$ rotation group. Indeed, they naturally form the standard $Spin(4)$ group for real fermions under considerations. Their algebras are isomorphic to  those of an $SO(4)$ group;

\begin{equation}
Spin(4)=SU(2) \otimes SU(2) \rightarrow SO(4).
\end{equation}
So when acting on real fermion fields, they naturally induce $SO(4)$ rotations in the Hamiltonian of {\em real fermion} fields.
In the later discussions, we will mainly focus on the algebraic aspect of this group rather than the topological ones and sometimes use $Spin(4)$ and $SO(4)$ interchangeably without further distinguishing them.
$Spin(n), n=4$ group naturally emerges in our current discussions of symmetry as spin groups are generic to real or charge neutral fermions.
They are also special limits of more general Clifford algebras defined for general Riemannian manifolds. We refer to more general discussions on relations between spin groups and Clifford algebras to Ref.\cite{CliffordA}.
From physics point of view, spin groups $Spin(n)$ can be thought as quantized version of classical groups $SO(n)$ as groups of $Spin(n)$ take into account essential quantum aspects of real fermion spin dynamics.

 

\section{Summary of the main results}

The main conclusions are that in gapless superfluids, generically there can be unexpected emergent $U(1)$ symmetries different from the conventional charge-$U(1)$ symmetries that are already spontaneously broken in superfluids (or superconductors). The results obtained below are inferred by the {\em EFT} introduced in the previous sections and symmetry structures discussed there. When applied to fermionic superfluids that already break the charge $U(1)$ symmetries or to superconductors, we find that there are naturally emerging $U(1)$ symmetries in infrared limits. And these emergent $U(1)$ symmetries imply a class of new conserved {\em charges} in a variety of gapless superfluids or superconductors which is the main subject of investigation in this article.

These emergent $U(1)$ symmetries can appear in the infrared limit even when real fermions are interacting (but when certain less relevant irrelevant interactions are suppressed). Moving into higher energy windows,
$U(1)$ symmetries usually are reduced to lower $Z_2$ symmetries. In the presence of strong interactions or generalized external fields when superfluids become gapped, 
the $U(1)$ symmetries are always broken while the Lorentz symmetry is still intact. 

We also find that emergent $U(1)$ symmetries are {\em always present} in gapless superfluids or superconductors even when the emergent Lorentz symmetry of $SO(3,1)$ is broken down to 
a lower $SO(3)$ or $SO(2)$ spin rotation symmetry. In the limits we have considered, all gapless states have an emergent $U(1)$ symmetry, disregarding the space-time symmetry.  It seems to suggest that emergent $U(1)$ symmetries are a 
characteristic of a broad class of gapless superfluids. All the $U(1)$ symmetry groups that appear in our studies are generated by a superposition of six linearly independent generators in an $Spin(4)$ group.

Finally, let us point out a close connection as well as differences between the fundamental or intrinsic charge-$U(1)$ symmetry in Weyl fermions and the emergent $U(1)$ symmetries in gapless superfluids (or superconductors) discussed here where the conventional charge $U(1)$ symmetry has been broken. For gapless stats with Lorentz symmetry, the mapping between these two classes of theories turn out to be a convenient starting point of discussions as illustrated in Appendices. 
Although mathematically there are closely related and indeed can be mapped into each other, {\em a single Weyl cone} is not a valid low energy sector for $3D$ lattice Weyl fermions with charge $U(1)$ symmetry, because of the well-known fermion doubling problem. It can only be a valid representation of single-cone $3D$ Weyl fermions living on the surface of a $4D$ bulk, a highly hypothetical situation. However, such a single Weyl cone turns out to be a valid representation of $3D$ bulk lattice fermions in superfluids, entirely due to the breaking of the usual charge $U(1)$ symmetry. Therefore, {\em EFT}s for gapless superfluids appear to be anomalous from the standard 3D bulk Weyl fermion points of view.

In this sense, the close relation between {\em EFT}s is purely at a level of theoretical abstraction rather than at a level of physical reality. See next section for more discussion. 
{\em EFT}s of $3D$ gapless superfluids can in fact be mapped into single-cone Weyl fermions that don't exist in 3D lattices because of fermion doubling.
Physically, all the different types of $U(1)$-{\em charges} that emerge in superfluids or superconductors as consequences of the emergent $U(1)$ symmetries are fundamentally distinct from the conventional particle number conservation in metals or insulators. 

And as stated before, $U(1)$ symmetries also emerge in the limit when the Lorentz symmetry is broken down to lower rotational symmetries in the presence of various condensation. In those cases, there are no longer explicit connections between Weyl fermions discussed in standard relativistic quantum field theories and gapless states that we are interested in.

\section{Emergent $U(1)$ symmetries and conserved charges in {\em EFTs}}

In Appendix A and B, we discuss how to identify an {\em EFT} for a particular system using a specific basis that defines the unique relation between real fermions and physical fermions (See also Eq.\ref{fixing}).
Namely, $\chi^T =(\chi_1,\chi_2,\chi_3,\chi_4)$ are {\em uniquely} defined as 

\begin{eqnarray}
&& \chi_{1} ({\bf r} )=\frac{1}{\sqrt{2}} [ \psi_{\uparrow} ({\bf r}) + \psi^\dagger_{\uparrow} ({\bf r})],  \nonumber \\
&& \chi_{2} ({\bf r} )=\frac{1}{\sqrt{2}} [ \psi_{\downarrow} ({\bf r}) + \psi^\dagger_{\downarrow} ({\bf r})], \nonumber \\
&& \chi_{3} ({\bf r} )=\frac{i}{\sqrt{2}} [ \psi_{\uparrow} ({\bf r}) - \psi^\dagger_{\uparrow} ({\bf r})],  \nonumber \\
&& \chi_{4} ({\bf r} )=\frac{i}{\sqrt{2}} [ \psi_{\downarrow} ({\bf r}) - \psi^\dagger_{\downarrow} ({\bf r})].
\label{fixing}
\end{eqnarray}
Here ${1,2,3,4}$ are indices for the real fermions, and $\uparrow, \downarrow$ are ones for spins or pseudo spins depending on microscopic starting points. $\psi^\dagger_{\uparrow,\downarrow}$, 
($\psi_{\uparrow,\downarrow}$) are
the creation (annihilation) operators of complex physical fermions. 

Using this specific basis of real fermions,
each physical system then will has a unique concrete {\em EFT} in the general form of Eq.(\ref{Eff}) but with definitive ${\bf \alpha, \beta}$ matrices and a definitive Lorentz group structure.
For each physical gapless states, there will be a unique emergent $U(1)$ symmetry or $U(1)$-charge that is conserved in the infrared limit. All the possible $U(1)$ symmetry groups
are generated by one of the six linearly independent $Spin(4)$ group generators or a superposition of them.

Furthermore, one can also show {\em EFT}s obtained in this specific way for different  physical systems can be connected with each other by rotations in the $Spin(4)$ group.
The mapping establishes an equivalence between a variety of $U(1)$ symmetries that appear in very different gapless superfluids. As illustrated in Appendices, they can all be mapped into the charge $U(1)$ symmetry of a hypothetical single-Weyl
cone in 3D. It is worth remarking again that physically, the single-Weyl-cone structure is forbidden in 3D lattices because of fermion doubling problem and in theory can only exit on the surface of a $4D$ lattice.
However, in fermionic superfluids (or superconductors), effective dynamics of a single Weyl cone do naturally appear in the gapless limit as a result of spontaneous charge-$U(1)$ symmetry breaking.

Now we are ready to focus on hidden $U(1)$-symmetries that can emerge in gapless superfluids or superconductors.
 At first sight, superfluids break the standard $U(1)$ global symmetries spontaneously. 
So the hidden $U(1)$ symmetries generally differ from the usually charge-$U(1)$ symmetry that leads to particle number or charge conservation. However, it turns out that the standard $U(1)$ charge-symmetry can be viewed as one of {\em six} more general linearly independent emergent $U(1)$ symmetries in gapless superconductors or superfluids.
In fact, the {\em charge} $U(1)$ symmetry group as well other more general hidden $U(1)$ symmetries are all embedded in an $Spin(4)$ group or an $SU(2) \otimes SU(2)$ group.  
They are represented by invariance of the Hamiltonians under one of the special $K$-type or $F$-type $SU(2)$ rotations and therefore we can name these hidden symmetries as $\tau$ or $\sigma$ symmetries in general.

The physical significance of emergent $U(1)$ symmetry depends on concrete states of physical fermions that are probed and measured, 
In this section we will carry out discussions using {\em EFTs} introduced in Appendix A and B and identify the six possible {\em linearly independent generators} of emergent $U(1)$ symmetries in in a broad class of gapless superfluids.
They form a complete set and all emergent $U(1)$ symmetry groups are generated by a linear superposition of these six indepedent charges. 
In the next section, we explore more practical physical consequences.
  
\subsection{$U_0(1)$ symmetry and chiral charge conservation}

To start with, we focus on 
 the simplest {\em EFT}, $H_{Weyl}$ in Eq.(\ref{Weyl}) that can be identified as an {\em EFT} of interacting Weyl fermions in the real fermion basis introduced above.
Eq.(\ref{Weyl}) has a {\em free} or non-interacting real fermions fixed point that is infrared stable in spatial dimensions higher than one or $d >1$. The dynamics of gapless fermions in the infrared limit turn out to be equivalent to  
simple {\em single} copy of non-interacting Weyl fermions as the partition functions of these two systems are identical.
The emergent symmetry discuss here is an asymptotic one even when interactions are included. We first exam the fixed point Hamiltonian with $g_{1,2}$ set to be zero.

The $U(1)$ global symmetry is generated by the operator

\begin{eqnarray}
Q_0=Q^\tau_2 &=&\frac{1}{2}  \int d{\bf r} q^\tau_2 ({\bf r}), U_2=\exp( i \frac{\phi}{2} Q_0); \nonumber \\
q^\tau_2({\bf r}) &=& \chi({\bf r}) \tau_y \otimes 1 {\chi({\bf r}}).
\end{eqnarray}
Hamiltonian $H_{Weyl}$ remains invariant under this $U(1)$ transformation and $Q_0$ is a conserved charge at the non-interacting fixed point when $g_1=g_2=0$.
This is an obvious result as $H_{Weyl}$ also coincides with a free complex fermion fixed point. For this reason, the model doesn't break the $U(1)$ global charge symmetry and the $U(1)$ charge, $Q_0$, shall be conserved.

The above $U(1)$ transformation is a special $F$-rotation around the axis ${\bf n} =(0,1,0)$, i.e. $F_2=\tau_y \otimes 1$ and therefore results in a rotation in $F_1-F_3$ plane (see Appendix.A). The other two charges of $F_1, F_3$ are rotated accordingly,

\begin{eqnarray}
Q^\tau_1&=& \frac{1}{2} \int d{\bf r} q^\tau_1({\bf r}),
q^\tau_1({\bf r})=\chi({\bf r}) \tau_x \otimes \sigma_y {\chi({\bf r}}); \nonumber \\
Q^\tau_3 &=& \frac{1}{2} \int d{\bf r}
q^\tau_3({\bf r}), q^\tau_3({\bf r})=
 \chi({\bf r}) \tau_z \otimes \sigma_y {\chi({\bf r}});
 \nonumber \\
Q^\tau_1&& \rightarrow Q^{'\tau}_1=\cos {\phi} Q^\tau_1 +\sin{\phi} Q^\tau_3,  \nonumber \\
Q^\tau_3 && \rightarrow Q^{'\tau}_3=-\sin {\phi} Q^\tau_1 +\cos{\phi} Q^\tau_3. 
\end{eqnarray}

In addition, under the time reversal transformation, 

\begin{equation}
Q^\tau_{1,2} \rightarrow Q^\tau_{1,2},  Q^\tau_3 \rightarrow -Q^\tau_3.
\end{equation}
Finally, as expected

\begin{eqnarray}
[ Q^\tau_i, Q^\tau_j ]=2i\epsilon_{ijk}Q_k^\tau, i,j,k=1,2,3.
\end{eqnarray}

This is a rather peculiar limit as physically a single copy of Weyl fermions with a given chirality are usually forbidden in a bulk $3D$ because of the well-known fermion double problems\cite{Nielson81}.
Such single-Weyl-cone phenomena usually only occur on a 3D surface of a 4D lattice so are hypothetical from physics point of view if one is interested in the physics in 3D bulk.

Nevertheless, dynamics of single copy of Weyl fermions can appear in an effective theory of the bulk of gapless superfluids (See below for more elaborated discussions). 
As shown in the Appendix B, {\em EFT}s of all the gapless superfluids discussed below are equivalent to $H_{Weyl}$ and its dual up to an $Spin(4)$ rotation.

Weyl fermions with single chirality as suggested in $H_{Weyl}$ surprisingly can appear as emergent  fermions in 3D topological superfluids. The corresponding {\em EFTs} can also have an infrared unstable strong coupling fixed point in spatial dimensions lower than three. At those fixed points, there is an additional emergent supersymmetry and states have higher symmetries than the non-interacting fixed point.
Beyond that point, $U(1)$ symmetry of $Q_0$ is broken spontaneously and states are gapped. In this case, they can form a topological superconducting state with time reversal symmetry if the symmetry is broken along the direction of
$Q^\tau_1$ or $F_1$ as $Q^\tau_1$ is even under the time reversal transformation. However, if the symmetry is not broken strictly along the direction of $Q^\tau_1$, then the T-symmetry can also be broken and the gapped states will be superconducting
without symmetry protected topological features.

$Q_0$ conservation here is obviously closely related to chiral symmetries discussed in quantum field theories and implies the conservation of Weyl fermions, either right handed or left handed.
However, there is a very important fundamental difference between the physics in superfluids and in the standard quantum field theory. In quantum field theories, fermions with both chiralities do appear coupled at certain ultraviolet scales and chiral anomalies induced by topological instantons eventually reduce the $U_L(1) \otimes U_R(1)$ to the simple $U_{L+R}(1)$ charge gauge symmetry\cite{t'Hooft76}. The right handed or left handed charges are not separately conserved as a result of chiral anomalies.

However, if Eq.(\ref{Weyl}) is taken as an effective field theory of gapless $p$-wave superfluids in a rotated basis (see Appendix B), only right handed  (or left handed) fermions with single chirality can emerge
because the underlying 3D topological superfluid state has odd parity under the parity transformation. That is the order parameter is odd under the parity transformation

\begin{eqnarray}
\Delta_{\alpha\beta} (-{\bf p})= -\Delta_{\alpha\beta}({\bf p}),\alpha,\beta=\uparrow, \downarrow.
\end{eqnarray}
The ground state transforms non-trivially under this parity transformation.
Therefore, in a given superfluid state where symmetries are broken spontaneously, only one chirality can emerge in effective field theories.
Fermions with opposite chirality only live in a different superfluid state or universe that is related to the one under consideration via a parity transformation.

Unless these two ground states can be connected by quantum tunnelling processes, dynamics in a given superfluid can only be related to either left-handed or right handed ones but not both.
In $(3+1)$D, $U(1)$  gauge monopoles are absent enforcing single chirality of real fermions. 
On the other hand, dynamics in two different superfluids related by parity transformation can both be represented by the single Weyl cone dynamics, left or right. Apart from that, they are identical.

For this reason, emergent chiral charges, as {\em emergent fermions}, can be conserved without usual anomalies extensively discussed in quantum field theories.
This emergent symmetry and conserved charges in the non-gauge models discussed above appear to be more robust than the chiral charge conservation that only appears at a classical level in quantum field theories of Yang-Mills gauge fields and in 
QED\cite{t'Hooft76,Adler69,Bell69}.

Finally, a generator of emergent symmetry always commutes with the Hamiltonian of {\em EFT}s, i.e. commutes with the non-interacting fixed point Hamiltonian that exhibits the Lorentz symmetry.
So the generator also commutes with both spin rotation subgroup generators, $S^{ij}$ and boost operators $S^{0i}$, which are basically constructed out of the free Hamiltonian (see discussions on $\alpha, \beta$ matrices Section II.).
As it commutes rather than {\em anti-commuting} with $S^{0i}$, such symmetry generators are not Lorentz invariant and do always break the Lorentz symmetry.

Before leaving this section, we want to point out that Weyl physics is also useful for discussions on helium-3 A phase\cite{Leggett75,Volovik03} where two nodal points emerge on two opposite sides of a Fermi surface.
In that case, two copies of two-component-complex fermions, or, eight real fermions are needed to describe dynamics. Each nodal point forms a representation of left and right Weyl fermions respectively as a result of fermion doubling.
From fermion doubling point of view, helium-3 A phase is a conventional state with expected numbers of fermi degrees of freedom while {\em EFT}s here do not have the issue of fermion doubling. 
That is {\em EFT}s discussed here and below only carry half of degrees of freedom in helium-3 A phase in low energy sectors so avoiding the fermion doubling problem.

\subsection{$U_{1,3}(1)$ symmetry and charge conservation of $Q_{1,3}$}

Now we are on the course to explore much less obvious emergent $U(1)$ symmetries in other more subtle cases presented in the previous section. 
The emergent symmetries in time-reversal invariant gapless $p$-wave superfluids in Eq.(\ref{TSF}) and time-reversal symmetry breaking gapless superfluids in Eq.(\ref{TSF1}) (see also Fig.\ref{EMS1})can be carried out in a similar way as we have done in the previous subsection.

In the {\em EFT} of time-reversal-invariant gapless superfluids in Eq.(\ref{TSF}), 

\begin{eqnarray}
&& Q^\tau_1 = \frac{1}{2} \int d{\bf r} q^\tau_1({\bf r}),  U_1=\exp( i \frac{\phi}{2} Q^\tau_1), \nonumber \\
&& q^\tau_1({\bf r})=\chi({\bf r}) \tau_x \otimes \sigma_y  {\chi({\bf r}})
\end{eqnarray}
defines a $U(1)$ transformation that leaves the {\em EFT} invariant.
Charge $Q_1$ is therefore conserved. 

In the {\em EFT} of time-reversal-symmetry breaking gapless superfluids in Eq.(\ref{TSF1}), 

\begin{eqnarray}
&&Q^\tau_3=\frac{1}{2} \int d{\bf r} q^\tau_3({\bf r}), U_3=\exp( i \frac{\phi}{2} Q^\tau_3), \nonumber \\
&& q^\tau_3({\bf r})= \chi({\bf r}) \tau_z \otimes \sigma_y  {\chi({\bf r}})
\end{eqnarray}
defines a $U(1)$ gauge transformation that leaves the {\em EFT} invariant.
Charge $Q^\tau_3$ is conserved.

$Q^\tau_{1,2,3}$ are three possible conserved charges of real fermions; their corresponding symmetry groups are embedded in an $SU(2)$ subgroup of $\tau$-type, dual to spin rotation $SU(2)$ subgroup of $S^{ij}$.
And for a given gapless state of superfluid, only one of them can be conserved. When this emergent $U(1)$ symmetry is further broken, superfluids become gapped and none of the charges above are conserved.

\subsection{$U_{1,2,3}$ symmetry and $\tau-\sigma$ duality}

Because of the $\tau-\sigma$ duality, there are three $\sigma$ charges defined below

\begin{eqnarray}
Q^\sigma_1 & =& \frac{1}{2} \int d{\bf r} \chi({\bf r}) \tau_y \otimes \sigma_x  {\chi({\bf r})}, U_1=\exp( i \frac{\phi}{2} Q^\sigma_1), \nonumber \\
Q^\sigma_2 &= & \frac{1}{2} \int d{\bf r} \chi({\bf r}) {I} \otimes \sigma_y  \chi({\bf r}), U_2=\exp( i \frac{\phi}{2} Q^\sigma_2), \nonumber \\
Q^\sigma_3 &=& \frac{1}{2} \int d{\bf r} \chi({\bf r}) \tau_y \otimes \sigma_z  \chi({\bf r}), U_3=\exp( i \frac{\phi}{2} Q^\sigma_3).\nonumber \\
\label{sigmac}
\end{eqnarray}
Three $U(1)$ transformations $U_{1,2,3}$ above leave {\em EFTs} in Eq.(\ref{DTS}),(\ref{DTS1}),(\ref{DW}), respectively, invariant. And so $Q^\sigma_{1,2,3}$ are conserved respectively.
Again if such a symmetry is broken, gapless states need to be fully gapped and there will be no more emergent $U(1)$ symmetries or conserved charges.

Just like $Q^\tau_{i}$, $Q^\sigma_i$, $i=1,2,3$ satisfy the usual $SU(2)$ algebras. Furthermore, they all commute with each of $Q^\tau_{i}$, $i=1,2,3$;

\begin{eqnarray}
[ Q^\sigma_i, Q^\sigma_j]=2i\epsilon_{ijk} Q^\sigma_k, [Q^\sigma_i, Q^\tau_j]=0, i,j,k=1,2,3.
\end{eqnarray}

\subsection{Emergent $U(1)$ symmetries: Effects of interactions}

So far, we have discussed emergent $U(1)$ symmetries exactly at a free fermion fixed point, the asymptotic theory.
Now we turn to the most subtle effects of interactions and intend to answer the question of whether these surprising symmetries are generic. 

In the presence of interactions, various $U(1)$ symmetries discussed above become much less obvious and in general there are no fundamental reasons why they shall be present. 
As we will see below, they can still emerge but only in the infrared limit where energy scales are low enough so that other {\em less relevant irrelevant} operators 
that we haven't included in Eq.(\ref{TSF}),(\ref{TSF1}),(\ref{Weyl}),(\ref{DTS}),(\ref{DTS1}),(\ref{DW}) play little role. However, once in an intermediate energy window where those muted interactions become important, $U(1)$ symmetries discussed above
are no longer valid and dynamics start substantially deviating from what we have described above.  

At first sight, {\em EFT}s with interactions $g_{1,2}$ appear to break $U(1)$ symmetries that are present in the non-interacting limit of $g_{1,2}=0$.
However, the key observation is that away from the non-interacting fixed point, these {\em most relevant irrelevant interaction operators} are actually invariant under those $U(1)$ rotations.
This is largely because there is only one single most relevant irrelevant 4-fermion operator  (i.e.without gradient operators) we can construct for four-component real fermions.
And so all interaction operators appearing in our {\em EFT}s are identical; i.e.

\begin{eqnarray}
H_{Ix} & =& H_{Iy}=H_{Iz} \nonumber \\
H_{Ix}&=& \int d{\bf r}
\chi^T \tau_x \otimes \sigma_y \chi \chi^T \tau_x\otimes \sigma_y \chi , \nonumber\\
H_{Iy} &=& \int d{\bf r}
\chi^T \tau_y \otimes I \chi \chi^T \tau_y \otimes I \chi , \nonumber\\
H_{Iz}&=&\int d{\bf r}
=\chi^T \tau_z \otimes \sigma_y \chi \chi^T \tau_z \otimes \sigma_y \chi 
\label{OP}
\end{eqnarray}
As far as only such interactions that are entirely local in $\chi$ fields are present in {\em EFT}s, $U(1)$ symmetries are still emergent just like in the non-interacting fixed point.

However, we do not expect these symmetries are present at higher energy scales when further moving away from the infrared fixed point and other {\em less relevant irrelevant interaction operators} play more important roles.
One such example are interaction operators of the form,

\begin{eqnarray}
H_{LR}&=&\int d{\bf r}
\chi^T \tau_z \otimes \sigma_x \nabla \chi  \chi^T \tau_z \otimes \sigma_x \nabla \chi  \nonumber \\
\label{GO}
\end{eqnarray}
where the subscript (LR) implies less relevant. Presence of such an interaction operator involving  gradient fields $(\chi \nabla \chi)^2$ invalidates the emergent $U(1)$ symmetry.
However, the scaling dimensions of such operators are higher than $H_{I}$ in Eq.(\ref{OP}), i.e.

\begin{eqnarray}
Dim[ H_I ] =d-1, Dim [ H_{LR}] =d+1,
\end{eqnarray} 
near the infrared stable fixed point. Hence, $U(1)$ symmetry always emerges as an asymptotic symmetry in the infrared limit.

Below, we summarize our main conclusions.
$U(1)$  symmetry become emergent {\em if}  one of the following conditions is met:

a) microscopic {\em UV} theories further indicate only a contact interaction with zero range appears at ultraviolet scales of {\em EFTs} to forbid the presence of operators of the form of $H_{LR}$ or alike in our {\em EFTs};

b) the $U(1)$ symmetry is also a fundamental one suggested by the intrinsic symmetry of the underlying physics;

c) in the infrared limit near the non-interacting fixed point when $H_{LR}$ are strongly suppressed; remaining interactions without gradient fields do exhibit the same $U(1)$ symmetry as the fixed point.

Both a) and b) are additional inputs to {\em EFT}s, rather than intrinsic of {\em EFT}s themselves. On the other hand generically, c) can always be satisfied as far as we lower the energy scales so to close to the infrared fixed point. 

For instance, if $H_{Weyl}$ is intended as a real Fermion representation of free Weyl fermions, then $U(1)$ symmetry itself is a fundamental global symmetry that {\em EFTs} shall also have and the emergent symmetry is the same as the fundamental symmetry.
The global gauge invariance enforces a $U(1)$ symmetry, or the condition b) is satisfied. And $U(1)$ symmetry is preserved by all allowed interaction operators, not surprisingly.
In this case, the $U(1)$ symmetry not only appears near the infrared fixed point but also in intermediate or higher energy scales and the corresponding $U(1)$ charge conservation becomes exact.

However, if $H_{Weyl}$ is intended as a representation of a gapless superfluid which already breaks the $U(1)$ global symmetry associated with charges, 
then the  $U(1)$ symmetry shown above is an emergent symmetry, either specifically for certain interactions
so that the condition a) is met, or alway appearing in the infrared limit as stated in c). 
In the following section, we will discuss  the generic case c) so that $U(1)$ symmetry does emerge in low energy sectors whenever {\em EFTs} are applicable.

\section{Emergent $U(1)$ symmetries and applications in gapless superfluids or superconductors}

Now we are going to apply these emergent symmetries and conserved charges and find out physical consequences in more specific systems.
We will focus on a few mostly commonly seen states although it can be easily generalized to many other possible states. Again in this section, we have assumed an infrared limit where interactions are fully captured by
the discussions in the previous section.
 
\subsection{Applications to Strong coupling limit of p-wave superfluids with $T$ reversal symmetries}

As stated before, {\em EFT} in Eq.(\ref{TSF}) can be directly applied to understand time-reversal-symmetric spinful superfluids or superconductors.
$\chi$ fields are related to physical complex fermion fields $\psi$ in a standard way,

\begin{eqnarray} 
\psi_{\alpha} ({\bf r}) =\frac{1}{\sqrt{2}} [\chi_{1\alpha} ({\bf r} )+i \chi_{2\alpha} ({\bf r})], \alpha=\uparrow, \downarrow.
\label{RC}
\end{eqnarray}

Using the above convention, we can identify Eq.(\ref{TSF}) as an {\em EFT} for a $p$-wave superfluid or superconductor.
Furthermore, each {\em EFT} representation of real fermions that is generated by $F$ or $K$ rotations discussed before shall be uniquely related a distinct physics system.

In Eq.(\ref{TSF}), the spin rotation generators $S^{ij}$ of the emergent Lorentz symmetry in this superfluid are defined by $K$-rotations in Eq.(\ref{Asym11}) or by $Q^\sigma_{1,2,3}$.
On the other hand, $Q^\tau_{1,2,3}$ are related to the generators of the dual group, $S^{ij}_D$ or $F$ rotations.

Three spin rotationally invariant $\tau$-generators or charges for $F$ rotations in terms of complex fermions are 

\begin{eqnarray}
Q^\tau_1 & =&\frac{i}{2}
 \int d{\bf r} [\Psi^T({\bf r}) \sigma_y {\Psi({\bf r})} -\Psi^\dagger({\bf r}) \sigma_y {{\Psi^\dagger}^T({\bf r})}] \nonumber \\
Q^\tau_2 &= & \frac{1}{2}  \int d{\bf r} [ \Psi^\dagger({\bf r}) {\Psi({\bf r})} - \Psi({\bf r}) {\Psi^\dagger({\bf r})} ] \nonumber \\
Q^\tau_3 &=& \frac{1}{2}
\int d{\bf r} [ \Psi^T({\bf r}) \sigma_y {\Psi({\bf r})}+\Psi^\dagger({\bf r}) \sigma_y {{\Psi^\dagger}^T({\bf r})}].
\end{eqnarray}
where $\Psi^T =(\psi_\uparrow, \psi_\downarrow)$.

In gapless $p$-wave T-invariant superfluids that naturally appear in strong coupling limits near topological phase transitions,
only $Q^\tau_1$ is conserved due to the emergent $U(1)$ symmetry.  This charge conservation becomes exact in the limit of contact interactions but for more generic interactions it is an emergent symmetry  
in the infrared limit.
When vacuum expectation values of $Q^\tau_2$ or $Q^\tau_3$ are non-zero, the state becomes fully gapped and
$Q^\tau_1$ charge is no longer conserved.

\subsection{Applications to Nodal phases in Superfluids}

Next, we look into the {\em EFT} in Eq.(\ref{DW}). Using the identification scheme for complex fermions above, we verify that the {\em EFT} can be for a $3D$ nodal phase physically induced by a strong magnetic field along the $y$-direction with a parity symmetry.
When the magnetic field is decreased, the nodal phase will undergo a quantum Lifshitz transition of free majorana class into a gapped superconductor\cite{Yang21}. At zero field, the gapped superconductor can be a topological one with time reversal symmetry. 

In this particular case of Eq.(\ref{DW}), $\sigma_y=-1$ means {\em  left-handed} (L)  nodal point while $\sigma_y=+1$ for {\em right handed} (R) nodal point and $\sigma_{x,y,z}$ acting on the pseudo spin space of LR. The emergent $U(1)$ symmetry in {\em EFTs} is due to 
the underlying azimuthal symmetry around the direction of $y$. 
Unlike in the previous subsection,
the spin rotation subgroup $S^{ij}$ of the emergent Lorentz symmetry in this superfluid turns out to be defined by $F$-rotations in Eq.(\ref{Asym11}) or $Q^\tau_{1,2,3}$.

On the other hand, $Q^\sigma_{1,2,3}$ are all rotationally invariant under the action of spin group $S^{ij}$ or in this case, $F$-rotations generated by $Q^\tau_{1,2,3}$.
In fact, they are directly related to the generators of the dual group, $S^{ij}_D$ or $K$ rotations and hence all commute with $Q^\tau_{1,2,3}$.
These three $U(1)$ charges of $\sigma$-type defined in Eq.(\ref{sigmac}) are now directly related to charges at left- and right-handed nodal points.
Introduce a pseudo spin representation for left and right fields $\psi_{L,R}$ and four-component real fermion $\chi$-fields,

\begin{eqnarray}
\chi^T_1({\bf r}) &=& \frac{1}{\sqrt{2}} [\Psi^T +\Psi^\dagger ],  \chi^T_2({\bf r})=- \frac{i}{\sqrt{2}} [\Psi^T -\Psi^\dagger]  
\nonumber \\
\Psi^T({\bf r}) &=& \frac{1}{{2}}  [(1,i) \psi_R + (1, -i) \psi_L],
\label{pseudo}
\end{eqnarray}
where we have defined pseudo spin real fermion fields as $\chi^T_{1}=(\chi_{1\uparrow}, \chi_{1\downarrow})$, $\chi^T_{2}=(\chi_{2\uparrow}, \chi_{2\downarrow})$.
Following Eq.(\ref{sigmac}),(\ref{pseudo}), we find that in terms of complex fermions,

\begin{eqnarray}
Q^\sigma_1 & =&  \int d{\bf r}  \Psi^\dagger ({\bf r}) \sigma_x {\Psi({\bf r})} ,
 \nonumber \\
Q^\sigma_2 &= &  \int d{\bf r} \Psi^\dagger({\bf r}) \sigma_y {\Psi({\bf r})} =Q_R-Q_L, \nonumber \\ 
Q^\sigma_3 &=&  \int d{\bf r}  \Psi^\dagger ({\bf r}) \sigma_z {\Psi({\bf r})}.\nonumber \\
\end{eqnarray}

More explicitly,

\begin{eqnarray}
Q^\sigma_1 & =& i \int d{\bf r} [ \psi^\dagger_L ({\bf r}) {\psi_R({\bf r})} - \psi^\dagger_R ({\bf r}) {\psi_L({\bf r})}],
 \nonumber \\
Q^\sigma_2 &= &  \int d{\bf r} [ \psi^\dagger_R({\bf r}) {\psi_R({\bf r})} - \psi^\dagger_L({\bf r}) {\psi_L({\bf r})}]=Q_R-Q_L, \nonumber \\ 
Q^\sigma_3 &=&\int d{\bf r} [ \psi^\dagger_L ({\bf r}) {\psi_R({\bf r})} +\psi^\dagger_R ({\bf r}) {\psi_L({\bf r})}].
\end{eqnarray}
where $Q_L=\int d{\bf r} \psi^\dagger_L\psi_L$ is the number of complex fermions at L-nodal point and $Q_R=\int d{\bf r} \psi^\dagger_R \psi_R $ the fermion charge at $R$-nodal point.


For a nodal phase, the emergent $U(1)$ symmetry leads to the conservation of $Q^\sigma_2$ or $Q_R-Q_L$ as the Hamiltonian in Eq.(\ref{DW}) remains invariant under the rotations of $Q^\sigma_2$.
This is a surprising result: superfluids break the conventional global $U(1)$ symmetry and so $Q^\tau_2=Q_L +Q_R$ is not conserved.
Nevertheless, the difference between $Q_L$ and $Q_R$ is conserved.

Microscopically,
one can also verify that  this conservation law becomes exact when there are no back scattering terms from two left particles to right ones or vice versa; this condition leads to the $U(1)$ symmetry of $Q^\sigma_2$.
Practically, this can be achieved by forbidding Umklapp processes.

\begin{figure}
\includegraphics[width=6cm]{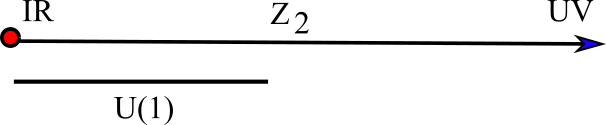}
\centering
\caption{ \footnotesize An $U(1)$ symmetry emerges in the infrared limit ({\em IR}) in an {\em EFT} that has a generic $Z_2$ symmetry up to an ultraviolet ({\em UV}) scale. 
Red dot indicates an infrared stable fixed point that dictates possible emergent symmetries in the limit of {\em IR}. The stability of the fixed point implies the robustness of emergent $U(1)$ symmetries.}
\label{EMS2}
\end{figure}

\subsection{$U(1)$ Gauge Symmetry}

So far we have treated the emergent symmetry as a global one and discovered various $U(1)$ charges carried by real fermions.
One can further extend the discussions and impose a $U(1)$ gauge symmetry of {\em EFT}s discussed before. A gauge symmetry leads to $U(1)$ charge-current conserved dynamics and 
below we briefly identify these currents.

We again focus on the limit where the spin group of the Lorentz transformation of $S^{ij}$ is an $SU(2)$ group isomorphically generated by $K$-type generators,
\begin{eqnarray}
K_1=\tau_y \otimes \sigma_x,  K_2=\tau_y \otimes \sigma_z, K_3=1 \otimes \sigma_y.
\end{eqnarray}
Concrete examples of this are the models in Eq.(\ref{TSF}), (\ref{TSF1}),(\ref{Weyl}).

In the case of Eq.(\ref{TSF}) that describes dynamics in gapless odd-parity superconductors, the emergent $U(1)$ gauge symmetry is generated by the following local transformation

\begin{eqnarray}
U_1 =\exp[-i\phi({\bf r}, t) Q_1^\tau], Q_1^\tau =\frac{1}{2} \int d{\bf r} \chi^T({\bf r}) \tau_x \otimes \sigma_y \chi({\bf r}).\nonumber \\
\end{eqnarray}

A gauge invariant {\em EFT} suggests a standard minimum coupling ({\em mc}) between the real fermion fields $\chi({\bf r})$ and emergent gauge fields $(A_0, {\bf A})$, i.e.
\begin{widetext}
\begin{eqnarray}
H_{mc} &=& \frac{1}{2} \int d{\bf r}  [A_0 \rho({\bf r}) + {\bf A} \cdot {\bf J}({\bf r}) ]; 
\rho({\bf r})  =\frac{1}{2} \chi^T({\bf r}) \tau_x \otimes \sigma_y \chi({\bf r}), \nonumber \\
{\bf J}_x({\bf r}) &= &\frac{1}{2} \chi^T({\bf r}) \tau_y \otimes \sigma_x \chi({\bf r}),
{\bf J}_y({\bf r})=  \frac{1}{2} \chi^T({\bf r}) I \otimes \sigma_y \chi({\bf r}),
{\bf J}_z({\bf r})=  \frac{1}{2} \chi^T({\bf r}) \tau_y \otimes \sigma_z \chi({\bf r}).
\label{current}
\end{eqnarray}
\end{widetext}

In terms complex fermions, these charge and current of emergent $U(1)$ gauge symmetry correspond to

\begin{widetext}
\begin{eqnarray}
\rho({\bf r}) =\frac{i}{2}[\Psi^T({\bf r}) \sigma_y {\psi({\bf r})} -\Psi^\dagger({\bf r}) \sigma_y {{\Psi^\dagger}^T({\bf r})}]; {\bf J}_x({\bf r})  = \Psi^\dagger ({\bf r}) \sigma_x {\Psi({\bf r})},
{\bf J}_y({\bf r})= \Psi^\dagger({\bf r}) \sigma_y {\Psi({\bf r})},
{\bf J}_z({\bf r})= \Psi^\dagger ({\bf r}) \sigma_z {\Psi({\bf r})} 
\label{current1}
\end{eqnarray}
\end{widetext}
where the $U(1)$ current density in this case is exactly a spin density field (up to a $1/2$ factor) while charge density is related to singlet condensation.

Real fermions are electrically charge neutral so do not respond to electric or magnetic fields.
However, the identifications in Eq.(\ref{current1}) imply possible interactions, analogous to {\em electric or magnetic} ones, between real fermions in gapless superfluids or superconductors and spin singlet condensates, as well as spin density structures. Potential applications and other implications will be further discussed in a separated paper that is under preparation.
More importantly, possible emergence or absence of gauge anomalies when gauging the global symmetries shall be further investigated.

\section{Emergent $Z_2$ symmetry in gapless superconductor or superfluids}

As stated in the previous section, a $U(1)$ symmetry can emerge in gapless superfluid or superconductors if one of the following four conditions is met:

i) exactly at an infrared stable non-interacting fixed point of {\em EFT}s in dimensions higher than $(1+1)D$;

ii) near the infrared fixed point where the less relevant irrelevant gradients fields are negligible and the remaining most relevant irrelevant interactions are invariant under $U(1)$ transformation;

iii) if interactions are strictly zero-range contact ones and the emergent symmetry is a result of this specific type interactions;

iv) if interactions in {\em EFT}s have to be subject to an additional $U(1)$ symmetry constraint as a consequence of intrinsic symmetries of microscopic {\em UV} model.

If neither of these requirements is satisfied, strictly speaking there can be no emergent $U(1)$ symmetries in quantum dynamics of {\em EFT}s. 
But are there remaining emergent symmetries for more general interactions in gapless superfluids? If there are, what are they?

Below we will discuss one generalization of the above {\em EFT}s to include not only interactions that is completely local in $\chi({\bf r})$-fields without gradient fields but also interactions that are non-local in space or time. 

Once interactions are generalized beyond the local fields of $\chi^4$-form,
instead, there can be a lower emergent symmetry that the $U(1)$ symmetry can be reduced to.
In other words, in more generic interacting systems, $U(1)$ symmetry only emerges at infrared energy scales when $H_{LR}$ are entirely renormalized to almost zero but at intermediate or higher energy scales, it breaks down to a smaller group.
In the example below, we will show explicitly that $U(1)$ symmetry evolves into $Z_2$ in higher energy scales.

\subsection{Generalized Interaction Model}
To better understand the origin and consequence of emergent $U(1)$ symmetry and limitation of the above analyses,
we further explore an extended  interacting real fermion model by explicitly introducing two real scalar fields $\phi_{1,2}=\phi_{1,2}^\dagger$,
We will see explicitly that $U(1)$ symmetry in {\em EFT}s discussed in the previous section appears to be an infrared-limit symmetry or an asymptotic symmetry of a more general quantum dynamics with lower symmetries.

To further carry out discussions, we illustrate our main point by working with an EFT  of the following form although one can easily arrive at the same conclusion by working with other rotated {\em EFT}s.

\begin{eqnarray}
H &=& H_0 +H_I ; \nonumber \\
H_0 &=& \frac{1}{2} \int d{\bf r}
[ \chi^T ({\bf r}) (\mathbb{I}  \otimes (\sigma_x  { i\nabla_x}+\sigma_z {i\nabla_z}) +\tau_y \otimes \sigma_y  {i\nabla y}) \chi ({\bf r}) 
 \nonumber \\
&+&\frac{1}{2} \sum_{i=1}^{2}\pi^2_i({\bf r}) + \nabla \phi_i ({\bf r}) \cdot \nabla \phi_i({\bf r})+ M^2_i\phi_i^2({\bf r})], \nonumber \\
H_I &=& \int d{\bf r}[ \sum_{i=1}^{2}g_{Yi} \phi_i({\bf r}) \chi^T({\bf r}) \tau_i \otimes \sigma_y  \chi({\bf r}) +g_4 (\sum_{i=1,2}\phi_i^2({\bf r}))^2 ], \nonumber \\
\label{SUSY3}
\end{eqnarray}
where $\tau_{1,2}=\tau_{x,z}$, $M_{1,2}$ are masses of real scalar fields $\phi_{1,2}$ respectively.
And we have further set the speeds of real scalar fields, $v_{1,2}=1$ so to have a desired emergent Lorentz symmetry.
Two real scalar fields $\phi_{1,2}$ and four-component majorana fields in Eq.(\ref{SUSY3}) are defined in a standard way,

\begin{eqnarray}
& &[\phi_i({\bf r}), \pi_j({\bf r}')] =i\delta_{ij} \delta({\bf r}-{\bf r'}), \nonumber \\
&& [\phi_i({\bf r}),\phi_j({\bf r'}]=[\pi_i({\bf r}),\pi_j({\bf r}')]=0, i=1,2; \nonumber \\
& &\chi^T= (\chi_{1\uparrow}, \chi_{1\downarrow}, \chi_{2\uparrow},\chi_{2\downarrow}).
\end{eqnarray}

In the special limit when $M_{1}=M_{2}$ and $g_{Y1}=g_{Y2}$, the model has a $U(1)$ symmetry and remains invariant if we make the following $U(1)$ transformation
\begin{eqnarray}
&\chi& \rightarrow \chi'=U(\phi)\chi=\exp(i \frac{\phi}{2}Q^\tau_2) \chi, \nonumber \\
&\Phi& \rightarrow \Phi'=U_B (\phi)\Phi=\exp(-i \phi) \Phi, \Phi=\phi_1+i\phi_2.
\end{eqnarray}

However, generically two masses $M_{1,2}$ are not equal so the {\em EFT} is not $U(1)$ invariant.
Instead, it is invariant under the following discrete transformation or a $\pi$-rotation,
\begin{eqnarray}
U(\pi)=\exp(i \frac{\pi}{2}Q^\tau_2), U_B=-1.  
\end{eqnarray}
The transformation $U(\pi)$ is purely real. 
Under this transformation,

\begin{eqnarray}
U^{\dagger}(\pi)  q^\tau_1({\bf r}) U(\pi) &=& -q^\tau_1({\bf r}) ,  U^{\dagger}(\pi)  Q^\tau_1 U(\pi) = -Q^\tau_1 ; \nonumber \\
U^{\dagger}(\pi)  q^\tau_3({\bf r})  U(\pi) &=& -q^\tau_3({\bf r}), 
U^{\dagger}(\pi)  Q^\tau_3 U(\pi) = -Q^\tau_3.
\end{eqnarray}

One can verify that for arbitrary mass ratio $M_1/M_2$, the Hamiltonian is indeed invariant under this transformation of $\chi$ fields combined with a reflection of $\phi$ fields, i.e.

\begin{eqnarray}
\chi_{1\alpha} &\rightarrow&  \chi_{2\alpha}, \chi_{2\alpha} \rightarrow -\chi_{1\alpha}, \alpha=\uparrow,\downarrow; \nonumber \\
\phi_{1} &\rightarrow & -\phi_1,   \pi_1 \rightarrow  -\pi_1, \phi_2 \rightarrow -\phi_2,   \pi_2 \rightarrow -\pi_2.
\label{RS}
\end{eqnarray}

In the limit when the running scale $\Lambda \ll M_{1,2}$, the massive bosons fields can be integrated out resulting in an {\em EFT} involving 4-fermion operators.
In the infrared limit $\Lambda /M_{1,2} \rightarrow 0$, the most relevant interactions in {\em EFT} are exactly of the local $\chi^4$-fields and a higher $U(1)$ symmetry emerges in this limit
as a result. However, when $\Lambda$ is comparable or larger than $M_{1,2}$, only $Z_2$ symmetry can be present in these intermediate scales.

\begin{figure}
\includegraphics[width=6cm]{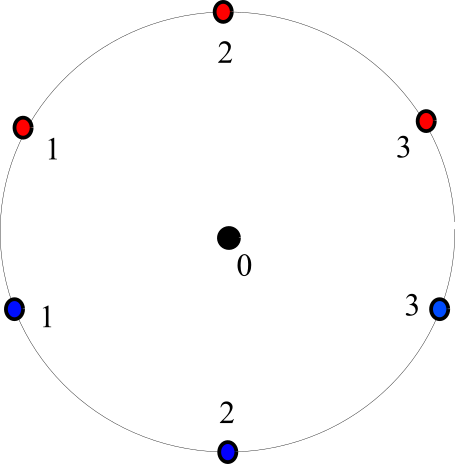}
\centering
\caption{ \footnotesize Breaking of emergent symmetries along different axis defined by generalized mass operators $\{ Q^\tau_{i}, i=1,2,3\}$ (top three in red) and its dual of $\{ Q^\sigma_{i}, i=1,2,3\}$ (bottom three in blue).
At each dot on the circle or along each direction, an expectation value of one of these six generators of group $Spin(4)=SU(2)\otimes SU(2)$ (that are isomorphic to $SO(4)$ ones) becomes non-zero. 
The gapless state labelled as $0$ in the centre has the highest emergent symmetries, with both a) Lorentz symmetry and b) $U(1)$ (or $Z_2$ symmetry). 
If the spin rotation subgroup of {\em EFT}s of state $0$ is defined by generators $\{ Q^\sigma_i, i=1,2,3 \}$ or $K$-type rotations, then only two of top three states $1,2,3$ are invariant under the Lorentz group $SO(3,1)$ but both break the emergent $U(1)$ or $Z_2$ symmetry. The third one is $SO(3)$ rotationally invariant but fully preserves the infrared $U(1)$ symmetry. On the other hand,
all the bottom three states break the $SO(3,1)$ Lorentz symmetry to an $SO(2)$ rotation one but again all display an emergent $U(1)$ symmetry. }
\label{EMS3}
\end{figure}

\subsection{$Z_2$ symmetry breaking}

Unlike the emergent $U(1)$ symmetry near the infrared free fixed point, this discrete $Z_2$ reflection symmetry appears to be a more generic emergent symmetry in gapless topological superfluids or superconductors with emergent Lorentz symmetries as it does
not require fine tuning of $M_{1,2}$. It can emerge at higher energies away from the infrared limit of $\Lambda \ll M_{1,2}$.
The emergent symmetries are again absent whenever the states are gapped and {\em EFT}s become massive.

In addition, Eq.(\ref{SUSY3}) can also exhibit a ${\cal Z}_2$-time reversal symmetry.
Further combined with this discrete time reversal symmetry, the model then is $Z_2 \otimes {\cal Z}_2$ symmetric.

Indeed, in the strong coupling limit of {\em EFT}s, the discrete $Z_2 \otimes {\mathcal Z}_2$ symmetry can be spontaneously broken resulting in massive real fermions.
This mechanism of mass generation is related to Gross-Neveu physics that had been well understood for Dirac fermions\cite{Gross74,Fei16}. For this to occur, one of the fields of $Q^\tau_1$ or $Q^\tau_3$ has to condense. That is 

\begin{eqnarray}
\langle Q^{\tau}_1 \rangle \neq 0, \mbox{ and/or}  \langle Q^{\tau}_3 \rangle \neq 0.
\end{eqnarray}
And if $Q^{\tau}_3$ is involved in the ${Z}_2$-reflection symmetry breaking, the time reversal symmetry is also broken as $Q^\tau_3$ is odd under the time reversal transformation.
However, if only condensation of $Q^\tau_1$ is involved in $Z_2$-reflection symmetry breaking, the time reversal symmetry is still intact as $Q^\tau_1$ is singlet under time reversal transformation.  

If $M_1=M_2=0$ and so $U(1)$-symmetry rather than $Z_2 \otimes {\cal Z}_2$ happens to be emergent at the strong coupling fixed point, a supersymmetry along with conformal symmetry shall also appear, similar to a strong coupling fixed point in a complex fermion nodal phase\cite{Lee07,Balents98,Fei16}. This feature had been recently applied to understand a strong coupling limit of topological quantum criticality with time reversal symmetry\cite{Zhou22}.

\section{Emergent $U(1)$ symmetry without Lorentz symmetry}

As mentioned in the previous section, there are six classes of $U(1)$ symmetries embedded in a $SU(2) \otimes SU(2)$ group generated by $Q^\tau_i, Q^\sigma_i$, $i=1,2,3$ that double covers 
an $SO(4)$ group.
For extended models with non-local interactions between $\chi$ fields, six $\pi$-rotations generated by these operators then define six different $Z_2$ symmetries for different systems in even high energy scales.

Below we list the dynamic consequence of breaking a $U(1)$  or $Z_2$ symmetry
starting from gapless Lorentz symmetric superfluids which have such a $U(1)$ or $Z_2$ symmetry. 
The effects of breaking emergent symmetry only depend on how these operators transform under the Lorentz group generated by $(S^{0i}, S^{ij})$, rotation group of $S^{ij}$, and a $Z_2$ group.

We now fix the spin rotation subgroup of the Lorentz group to be $\sigma$-type. 
More explicitly, $s_i=\frac{1}{2}\epsilon_{ijk} S^{jk}$, $i,j,k=x,y,z$ are related to $Q^\sigma_i$, $i=1,2,3$. 
Then, the dual of $S^{ij}$, $S^{ij}_D$ are associated to $Q^\tau_i$, $i,j=1,2,3$.

\begin{eqnarray}
\epsilon_{ijk} S^{jk}=Q^\sigma_i,  
\epsilon_{ijk}S^{jk}_D =Q^\tau_i; i=x,y,z.
\label{RD}
\end{eqnarray}

a) Only two out of three $Q^\tau_i$, $i=1,2,3$ are Lorentz invariant.  For {\em EFT}s in Eq.(\ref{TSF}), they are $Q^\tau_{1,3}$; when $\langle Q^\tau_i \rangle \neq 0$, $i=1,3$, a mass gap opens up but with full Lorentz symmetry of $SO(3,1)$.
However, $Z_2$ symmetry is alway broken. (See Fig.\ref{EMS3}) 
  
 b) The third $Q^\tau_i$ is rotation invariant but not Lorentz invariant. This is also the generator for $U(1)$ emergent symmetry group. For {\em EFT}s in Eq.(\ref{TSF}), this operator is $Q^\tau_1$.
 When $\langle Q^\tau_1 \rangle \neq 0$, a spherical Fermi surface emerges while the state remains gapless. The Lorentz symmetry of $SO(3,1)$ is broken but the subgroup $SO(3)$ rotation symmetry remains. In addition, 
$Z_2$ or $U(1)$ symmetry in the infrared remains unbroken.

 c) Three $Q^\sigma_i$, $i=1,2,3$, that are dual to $Q^\tau_i, i=1,2,3$, are also the generators of spin rotation subgroup of $SO(3,1)$. So they not only fully break the Lorentz symmetry $SO(3,1)$, and also break the subgroup rotation symmetry $SO(3)$. When $\langle Q^\sigma_i \rangle \neq 0$, $i=1,2,3$, gapless real fermions appear at two nodal points
 along one of the $x,y,z$ directions, i.e. a nodal phase emerges. 
Furthermore, $U(1)$ symmetry remains as $Q^\sigma_i$, $i=x,y,z$ are all singlets of $\tau$-symmetry group which defines the emergent $U(1)$ symmetry group in the current discussion. (See Fig.\ref{EMS3})

In this particular case, although the Lorentz symmetry (of $\sigma$ type) is broken, there can be an emergent Lorentz symmetry of $\tau$-type reappearing in the reconstructed {\em EFT}.  That is, in contrast to Eq.(\ref{RD}) 
 
\begin{eqnarray}
\epsilon_{ijk} S^{jk}=Q^\tau_i,  
\epsilon_{ijk}S^{jk}_D =Q^\sigma_i; i=x,y,z.
\label{RD1}
\end{eqnarray}

Overall, non-zero expectation values of these operators can be either due to applied external fields that explicitly break these symmetries or due to strong interactions that break these symmetries spontaneously. Either way, gapless superconducting states will become either fully gapped states with Lorentz symmetry but breaking the emergent $Z_2$ or infrared $U(1)$ symmetry or other gapless states that break the full Lorentz symmetry but preserve the emergent infrared $U(1)$ symmetry.
Note that all the gapless states under our considerations, with or without Lorentz symmetry, have emergent $Z_2$ or even $U(1)$ symmetry in the infrared limit.

\begin{figure}
\includegraphics[width=6cm]{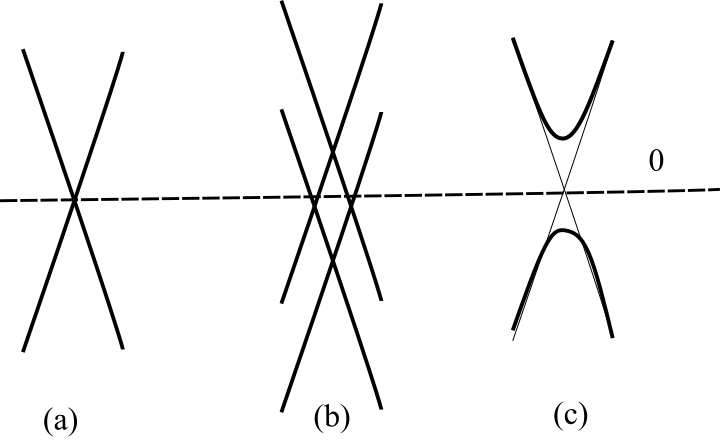}
\centering
\caption{ \footnotesize Schematic of dispersion relations in different gapless superfluids or superconductors with emergent $U(1)$ symmetries. a) Gapless states with highest symmetries, both  $SO(3,1)$ Lorentz symmetry and an emergent $U(1)$ symmetry (corresponding to state 0 in the centre of Fig.\ref{EMS3}); b) Gapless states with either $SO(3)$ or $SO(2)$ rotation symmetries but no full Lorentz symmetry due to condensation of one of Lorentz non-invariant charges in $\{ Q^\tau_{i}, Q^\sigma_i\}$, $i=1,2,3$. All these states have the same emergent $U(1)$ symmetry as the parent state in a);  c) Gapped states that break emergent $U(1)$ symmetries in a) but with full $SO(3,1)$ Lorentz symmetry. 
The dashed horizontal line indicates zero energy and the spectrum below the line is related to the above via a standard charge conjugation transformation in superfluids.
 }
\label{EMS4}
\end{figure}

\section{Conclusion}

In conclusion, we have investigated various emergent $U(1)$ or $Z_2$ symmetries that appear at different energy scales in gapless superconductors or superfluids and studied relations between them.
We show that emergent $U(1)$ symmetries in these gapless states can be conveniently characterized by $S^{ij}_D$, a group dual to $S^{ij}$, the spinor rotation subgroup of the emergent Lorentz symmetry group. 
Generators of symmetry groups are always singlets under the action of $S^{ij}$ of the rotation subgroup but are not invariant under the Lorentz group.
They can appear either in form of $U(1)$ symmetry in infrared limits or as a $Z_2$ symmetry in even intermediate energy scales. 

All the $U(1)$ symmetries are embedded in an $Spin(4)=SU(2)\otimes SU(2)$ group that double covers an $SO(4)$ group. $Spin(4)$ group has been identified as a product of an $SU(2)$ Lorentz spin rotation subgroup generated by $S^{ij}$, $i\neq j=x,y,z$, 
which is associated with $F$-(or $K$-)rotations and its dual
$SU(2)$ rotation group induced by $S^{ij}_D$, $i\neq j=x,y,z$ that is associated with $K$- (or $F$-)rotations. For real fermions, $K$ and $F$-rotationss or $SU_{K,F}(2)$ transformations are generated by {\em purely imaginary} hermitian operators that are algebraically defined in Eqs.(\ref{Asym11}),(\ref{KF}).
And all $U(1)$ emergent symmetry groups are shown to be spin rotation singlets under the spin rotation of $S^{ij}$. 

When the symmetry is broken spontaneously due to strong interactions,
the states become fully gapped while preserving the Lorentz symmetry. However, these emergent $U(1)$ global symmetries always remain if strong interactions break emergent Lorentz  $SO(3,1)$ symmetries spontaneously and 
only result in other gapless states with much lower symmetries such as $SO(3)$ or $ SO(2)$ rotational symmetries. As an open question, it is interesting to see what happens to emergent symmetries and whether emergent symmetry groups become bigger when the degrees of freedom of real fermions, or the central charges further increase to $4N$ with $N > 1$.   

There had also been quite a few intensive efforts to generalize the concept of gapped topological states or phases put forwarded before to gapless limits\cite{Kestner11,Fidkowski11,Sau11,Cheng11,Kraus13,Ortiz14,Ruhman15,Iemini15,Keselman15,Kainaris15,Montorsi17,Wang17,Kane17,Scaffidi17,Guther17,Verresen18,Jiang18,Verresen19,Thorngren21}.
And the vast majority of these efforts have been mainly on one-dimensional states. One interesting question that had been asked recently in Ref.\cite{Thorngren21}
is whether there are intrinsic topological states that can only be defined in the gapless limit and there are no gapped counter-parts. 

The main efforts in this article are to illustrate unique emergent symmetries in high dimensional gapless states. These may be important observations to future studies of gapless topological states and orders.
Broadly speaking, it remains to be fully understood whether these or other unique emergent symmetries in gapless states that always appear to be broken
in the gapped limit are essential for general understanding of topological physics\cite{Chen10,Chen12,Chen13,Wen17}.
The other interesting issue is possible local gauge symmetries associated with the emergent global symmetries discussed so far \cite{Witten18}. If we enforce such a local gauge symmetry, real fermions considered above can be further interacting with 
emergent dynamic $U(1)$ gauge fields or $Z_2$ gauge fields. These gauge fields can be in either weakly interacting phases or strong coupling confining phases.
In general, between these two phases there can even be highly symmetric quantum critical points with emergent gauge fields, similar to effective field theories discussed recently in Ref. \cite{Bi19,Ji20}. Perhaps an important question is under which conditions, such gauge fields and their interactions become practically relevant in studies of practical gapless superconductors or superfluids and what are potential implications.

The author wants to thank Z.-C. Gu, G. W. Semenoff, X. G. Wen and F. Yang for discussions on symmetries in gapless states.
This project is in part supported by NSERC, Canada under a Discovery grant under RGPIN-2020-07070.

\appendix

\section{Unitary Transformation of Real Fermions and $\tau-\sigma$ duality}

\subsection{Unitary Transformation}

Readers can skip this section and move to Sec.V if they are not interested in details of algebraical structures of the symmetry groups.

Unitary transformation of real fermions needs to leave fermions in a real representation. That requires that all unitary rotations be implemented in a real representation and so effectively become orthogonal transformations.
For that purpose, we can only utilize {\em pure imaginary, antisymmetry} generators. The only generators fall into this class are the ones specified as $K_{i}$ and $F_i$, $i=x,y,z$ in Eq.(\ref{Asym11}), (\ref{KF}) which are indeed isomorphic to
$SO(4)$ group algebras and generate the entire $Spin(4)$ group.

To highlight the structure of $Spin(4)$ group, we study the general rotations of real fermion fields induced by $SU_K(2)$ or  K-rotations and $SU_F(2)$ or F-rotations in $Spin(4)$ group. By construction, each group element in $Spin(4)$ can be specified as $[U_K, U_F]$, a pair of
$K$-rotation and $F$-rotation. 
Just like in standard constructions of $SU(2)$ rotations, we specify each of these $SU(2)$ rotations by  Euler angles $({\bf n}, {\phi})$,  i.e. a rotation axis ${\bf n}=(n_x, n_y, n_z)$ 
(${\bf n} $ is a unit vector) and a rotation angle $\phi \in [0,2\pi]$.
Each $Spin(4)$ group element is thus specified by two sets of Euler angles, one for $K$-group, i.e.$({\bf n}_K, \phi_K)$; the other for $F$-group, i.e. $({\bf n}_F, \phi_F)$. 

Under the action of $Spin(4)$ group, real fermions can transform accordingly. In our case, transformations of real fermions are by multiplication of the pair of $SU(2)$ rotations in $[U_K, U_F]$ of $Spin(4)$ group. That is 

\begin{eqnarray}
&& \chi \rightarrow U_T ({\bf n}_K, \phi_K; {\bf n}_F,\phi_F) \chi \nonumber \\ 
&&U_T ({\bf n}_K, \phi_k; {\bf n}_F,\phi_F)=U_K({\bf n}_K, \phi_K) \cdot U_F ({\bf n}_F,\phi_F)
\end{eqnarray}
where the subindices $K,F$ refer to $SU(2)$ rotations of $K$-type and $F$-type respectively.
In presenting this result, we have taken into account that each of three generators $K_i$,$i=x,y,z$ in the $SU(2)$ group of $K$ rotations  commutes with each of $F_i$, $i=x,y,z$ in the $SU(2)$ group of $F$ rotations. So, $U_T$ is simply a product of
two $SU(2)$ transformations, $U_K$ and $U_F$ in $Spin(4)$ group. Explicit structures of $U_{F,K}$ will be presented in the next section.

It is important to notice that $SU_K(2)\otimes SU_F(2)$, the product of $SU(2)$ groups of $K$-type $U_K$ and $F$-type $U_F$ are isomorphic to $S^3 \otimes S^3$. Each $S^3$ can be conveniently defined by a set of hyperspherical coordinates,
 
\begin{eqnarray}
(\cos\frac{\phi}{2}, \sin\frac{\phi}{2}{\bf n})
\end{eqnarray}
where $\phi \in [0, 2\pi]$ and ${\bf n}$ again is a three dimensional unit vector projecting out an $S^2$. 

An inversion in a single $S^3$ corresponds to $({\bf n}, \phi) \rightarrow (-{\bf n}, 2\pi-\phi)$.  Applying the standard $SU(2)$ algebras, one can verify explicitly that under such an inversion, indeed

\begin{eqnarray}
U_F(- {\bf n}_F, 2\pi-\phi_F) &=& - U_F({\bf n}_F, \phi_F);  \nonumber \\
U_K(- {\bf n}_K, 2\pi-\phi_K) &=& - U_K({\bf n}_K, \phi_K).
\end{eqnarray}    

Thus, under an inversion in an $S^3 \otimes S^3$, both $U_F$ and $U_K$ acquire minus signs. That is each element of $Spin(4)$ group acquires a minus sign under inversion; and is mapped into minus of itself, i.e. $[ U_K, U_F]  \rightarrow  [-U_K, -U_F]$. This however leaves the bilinear structure in $U_T$ invariant under such an inversion, although each of $U_{F,K}$ is not.
That is 

\begin{eqnarray}
U_T (-{\bf n}_K, 2\pi-\phi_K; -{\bf n}_F, 2\pi-\phi_F) =U_T ({\bf n}_K, \phi_K; {\bf n}_F, \phi_F). \nonumber \\
\label{DC}
\end{eqnarray}
A pair of inverted points in $S^3 \otimes S^3$ therefore lead to the same orthogonal rotation of real fermions. 
Eq.(\ref{DC}) indicates a double covering of $SO(4)$ by $S^3 \otimes S^3$ and hence by the spin group $Spin(4)=SU(2) \otimes SU(2)$ that is isomorphic to $S^3 \otimes S^3$.

In addition, spin rotations of $S^{ij}$ can be covered by an $SU(2)$ group; and $S^{ij}$ are also antisymmetry and hermitian. This indicates that they shall also be in one of the two $SU(2)$ groups, either $K$-group or
$F$-group in Eqs.\ref{Asym11},\ref{KF}. Whatever it is, it remains invariant under the actions of the other group.  

For instance, if $S^{ij}$, $i\neq j=x,y,z$ are represented by $K$-group generators, they are invariant under any action of $F$-group, and vice versa. 
However, under the actions of the same group of $S^{ij}$ or $K$-group actions, $S^{ij}$ can be further rotated and are not invariant. 
On the other hand, $S^{0i}$, $i=x,y,z$ are not invariant under actions of either group, $K$- or $F$-group.

\subsection{$\tau-\sigma$ Duality}
One shall notice that $K$- and $F$-rotations are related by a $\tau-\sigma$ duality,

\begin{eqnarray}
F_i \leftrightarrow K_i \mbox{  when $ \tau_i\leftrightarrow \sigma_i$.}
\end{eqnarray}
Therefore, generally speaking,  $Spin(4)=SU(2)_F \otimes SU(2)_K$ is simply generated by $S^{ij}$ and its $\tau-\sigma$ dual, whichever $S^{ij}$ are.
We name the $\tau-\sigma$ dual of $S^{ij}$ defined in Eq.(\ref{S}) as $S^{ij}_D$. $S^{ij}$ are also purely imaginary and antisymmetric and hermitian.  

\begin{eqnarray}
S^{ij} \leftrightarrow S^{ij}_D \mbox{  when $ \tau_i\leftrightarrow \sigma_i$.}
\end{eqnarray}
This simple observation, the invariance of $S^{ij}$ under its dual $S^{ij}_D$, either $K$-group or $F$-group,  suggests a way to classify {\em EFTs} and identify them with different physics reality depending on the structure of 
$S^{ij}$. 

$\tau$-symmetry below is assigned to {\em EFTs} where their Lorentz rotation subgroup of $S^{ij}$ is identified as $K$-rotations 
and remains invariant under the actions of its dual $S^{ij}_D$ or $F$-group. $\sigma$-symmetry is assigned to {\em EFTs} where their Lorentz rotation subgroup of $S^{ij}$ is identified as $F$-rotations and
remains invariant under the actions of its dual $S^{ij}_D$, which now is $K$-rotations.
$S^{ij}$ together with its $\tau-\sigma$ dual, $S^{ij}_D$, form the algebra group of $Spin(4)$ and generate the $Spin(4)$ group.

For each class, we can identify a parent Hamiltonian and generate the rest members in the class by the dual of $S^{ij}$, $S^{ij}_D$ which can be either $K$-group or $F$-group, whichever is the dual of $S^{ij}$. Within each class, {\em EFTs} have the same representation for the Lorentz subgroup group 
$S^{ij}$ up to a rotation induced by $S^{ij}$ itself; and we do not discuss such a trivial spin rotations generated by $S^{ij}$ in this article.

 The action of $S^{i,j}_D$, the dual of $S^{ij}$, on $S^{ij}$  is trivial as $S^{ij}$ is invariant under its dual $S^{ij}_D$; it leaves $S^{ij}$ unchanged. However, $S^{ij}_D$ acts on $S^{0i}$ or $\alpha_i$ non-trivially. It can generate other members of {\em EFTs} with the same $S^{ij}$ rotation group and they correspond to different physical systems. On the other hand, because different real fermion representations, or Hamiltonians are related by simple rotations, dynamics and the partition functions shall be obviously identical.
This is the focus of this article, to explore the classes of {\em EFT}s that form a group representation of the dual of the rotation group $S^{ij}$, i.e. the $SU(2)$ subgroup in $Spin(4)$ generated by $S^{ij}_D$.

\subsection{Generalized mass Operators}

All the bilinear operators that are even in momentum and hence can be non-vanishing in the limit of zero momentum are relevant operators or in short we call them {\em mass operators}. 
All  {\em mass} terms for real fermions are represented by anti-symmetric hermitian matrices, they have to be one of $K$- or $F$-rotation generators coinciding with the algebraic group of $Spin(4)$.
Two of them  are Lorentz invariant that are usually studied in quantum field theories but the rest four are not; further these four don't not lead to a mass gap in the spectrum unlike the other two nevertheless we simple call them {\em mass operators} in this article. Also as $K$-rotation and $F$-rotation generators are mutually commuting, all mass terms associated with $K$ generators are invariant under $F$-rotations and vice versa.

As we will be mainly interested in spatially rotationally invariant states and three rotation generators defined by $S^{ij}$ are either associated with $K$- or $F$-generators, the mass operators have to be identified as the dual of $S^{ij}$, that is $S^{ij}_D$ that can be either  $F$- or $K$-generators. 

The maximum number of mass terms allowed by a given gapless state with
$\alpha_i, i=x,y,z$ in  {\em EFT}s already identified is therefore {\em three}. Furthermore, we also find only two of these three rotationally invariant mass operators anti-commute with Lorentz boost operators $S^{0i}, i=1,2,3$, and
 are Lorentz invariant.

The third one that commutes with  $S^{0i}$ or $\alpha_i$, $i=x,y,z$ explicitly breaks Lorentz symmetry. On the other hand, it represents an emergent or hidden $U(1)$ symmetry in gapless states we will focus on.

The emergent $U(1)$ symmetry can be broken whenever one of the other two mass operators in $S^{ij}_D$, or both, condense and develop finite expectation values in the ground states while the Lorentz symmetry is preserved.
 
 At last, if the third mass operator which is Lorentz non-invariant condenses with non-zero expectation values, the Lorentz symmetry is broken; typically, there shall be an emergent Fermi surface in gapless states.
 
Finally, one can also have mass terms that further break $SO(3)$ rotation symmetries, in addition to breaking the Lorentz symmetry. This is actually a path leading to nodal phases.
However, in this case, a new Lorentz symmetry can re-emerge replacing the original Lorentz group; the mechanism is similar to the emergence of relativistic physics in $(1+1)$D due to emergence of Fermi points at fermion  surfaces.

\subsection{Identifying an {\em EFT} with a Physical System }

Real fermions are emergent particles in physical systems instead of fundamental ones. The relations between physical complex fermions and real fermions are usually simple in one
particular representation than other ones. So although dynamically different systems can be equivalent, when it comes to physical interpretations or predictions, we always prefer one with the simplest relation between physical complex fermions and
real fermions. If we specify real fermions $\chi$ directly in terms of complex ones $\psi$ for a given physical system and only work with a particular choice of real fermion fields, then {\em EFT} Hamiltonian in term of those real fermion
is entirely fixed. Further rotated {\em EFT}s involve $\chi$ fields defined in different ways in terms of complex fermions.  A popular choice is to identify

\begin{eqnarray}
&& \chi_{1} ({\bf r} )=\frac{1}{\sqrt{2}} [ \psi_{\uparrow} ({\bf r}) + \psi^\dagger_{\uparrow} ({\bf r})],  \nonumber \\
&& \chi_{2} ({\bf r} )=\frac{1}{\sqrt{2}} [ \psi_{\downarrow} ({\bf r}) + \psi^\dagger_{\downarrow} ({\bf r})], \nonumber \\
&& \chi_{3} ({\bf r} )=\frac{i}{\sqrt{2}} [ \psi_{\uparrow} ({\bf r}) - \psi^\dagger_{\uparrow} ({\bf r})],  \nonumber \\
&& \chi_{4} ({\bf r} )=\frac{i}{\sqrt{2}} [ \psi_{\downarrow} ({\bf r}) - \psi^\dagger_{\downarrow} ({\bf r})].
\label{fixing}
\end{eqnarray}
Here ${1,2,3,4}$ are indices for the real fermions, and $\uparrow, \downarrow$ are ones for spins or pseudo spins depending on microscopic starting points.

For this reason if we only work with this particular choice of $\chi$ fields or effectively fix an $Spin(4)$ {\em gauge}, every Hamiltonian in the two classes discussed above corresponds to one single physical reality. This is the point we will be taking in this article. Every physical system has a specific {\em EFT}. As each can be further related to other Hamiltonians in the same family if one performs a {\em purely real} unitary rotation of $\chi$, dynamically speaking, different physical systems can be mapped into other ones by redefining $\chi$ fields. And we will be exploring the relations between emergent symmetries in different states using the $K$-group and $F$-group.

\section{Mapping between real fermions}

\subsection{$F$-rotations and mapping}

We first focus on a class of {\em EFTs} where $SU(2)$ rotation subgroup of $S^{ij}$ in Lorentz group $SO(3,1)$ is given by $K$-rotations in $Spin(4)$ group.
We will study the members generated by the dual group $S^{ij}_D$ which in this case is $SU_F(2)$ subgroup in $Spin(4)$. All $F$-rotations leave $S^{ij}$ and their rotation group $SU_K(2)$ or $K$-rotations invariant; all {\em EFT}s
here have the same structures of $S^{ij}$, $i\neq j=x,y,z$.

We start with the most well known model for $p$-wave superconductors or superfluids in a strong coupling limit,
an {\em EFT}  for topological quantum critical points (TQCPs) with time reversal-symmetry.
\begin{widetext}
\begin{eqnarray}
H_{TS} =\frac{1}{2} \int d{\bf r} [ \chi^T({\bf r}) (\tau_z \otimes (\sigma_x {i\nabla_z} -\sigma_z {i\nabla_x}) +\tau_x \otimes \mathbb{I} {i\nabla_y}) \chi ({\bf r}) 
+ g_1 \chi^T \tau_y \chi \chi^T  \tau_y \chi +g_2 \chi^T  \tau_z\otimes \sigma_y \chi  \chi^T \tau_z\otimes \sigma_y  \chi ] \nonumber \\
\label{TSF}
\end{eqnarray}
\end{widetext}
where $\chi^T= (\chi_{1\uparrow}, \chi_{1\downarrow}, \chi_{2\uparrow},\chi_{2\downarrow})$.
$\chi$ are {\em real} fermions defined by the following standard algebra,

\begin{eqnarray}
&& \chi^\dagger_\alpha ({\bf r})=\chi_\alpha({\bf r}), \{ \chi_\alpha({\bf r}), \chi_\beta({\bf r}) \}=\delta_{\alpha,\beta}, \nonumber \\
&&\alpha,\beta=1\uparrow, 1\downarrow, 2\uparrow, 2\downarrow.
\end{eqnarray}
[ Here we have renamed $\chi_{1,2,3,4}$ in terms of $\chi_{1\uparrow, 1\downarrow,2\uparrow,2\downarrow}$ to illustrate the spin structures explicitly ]. 

The {\em EFT} in Eq.(\ref{TSF}) is a minimum representation for TQCPs\cite{Yang21,Zhou22} with spinful time reversal symmetry and emergent Lorentz symmetry. 
It can further have emergent supersymmetry in strong coupling limits. 
We now exam the effective field theories or {\em EFT} that we can further obtain via $SU(2)$ rotations generated by $F$-group in Eq.(\ref{KF}) and the $\tau$-symmetry associated with it.
The general structure for a rotation along the direction of ${\bf n}=(n_x, n_y, n_z)$ with an angle $\phi$ (following the right hand rule) is

\begin{eqnarray}
U_F({\bf n}, \phi) &= &\cos\frac{\phi}{2} -i\sin\frac{\phi}{2} (  n_x \tau_x \otimes \sigma_y +n_y \tau_y +n_z \tau_z \otimes \sigma_y); \nonumber \\
U_F({\bf n}, \phi) & = &U_F^{*}({\bf n}, \phi), U_F^{-1} ({\bf n}, \phi)=U^{T}_F({\bf n}, \phi).
\label{ROT}
\end{eqnarray} 
Note for $F$ rotations, the range of $\phi$ shall be  set as
\begin{equation}
2\pi \geq \phi \geq 0.
\end{equation}

As expected, $U_F({\bf n}, \phi)$ defines a simply connected three-sphere  manifold $S^3$. Specially,

\begin{eqnarray}
U_F(- {\bf n} ,2\pi- \phi)= -U_F({\bf n}, \phi)
\end{eqnarray}
which effective defines a pair of diagonally opposite points in an $S^3$.
On the other hand, $U_F({\bf n}, \phi)$ and $-U_F({\bf n}, \phi)$ obviously results in the identical rotations in the Hamiltonian manifold.
Inclusion of the other three $S^3$ defined by $U_K$ that is the dual of $U_F$ effectively allows a double coverage of an $SO(4)$ group by an $Spin(4)$ group, similar to a universal coverage of $SO(3)$ by a quantum spin group $Spin(3)=SU(2)$.

Below are two examples where ${\tilde H}$ in {\em EFT}s after $F$-rotations are transformed into the Hamiltonians in other superconducting states or superfluids when we apply the same identification in Eq.(\ref{fixing}) to the rotated real fermions.
Hence we show that interaction dynamics are completely identical as they are given by the same partition functions.

Under such a purely {\em real} rotation, 

\begin{eqnarray}
 \chi({\bf r})  \rightarrow {\tilde \chi}({\bf r}) =U_F \chi({\bf r}), H [\{ \chi({\bf r}) \}]  \rightarrow  { H} ={\tilde H} [\{ {\tilde \chi}({\bf r}) \} ]
 \nonumber \\
\end{eqnarray}
where fermions remain to be real, i.e.${\tilde \chi}^{T}({\bf r})={\tilde \chi}^\dagger({\bf r})$.

{\em A1: TQCP model with time-reversal-symmetry broken}

First we consider a rotation where

\begin{eqnarray}
{\bf n}=\frac{1}{\sqrt{3}}(1,1,1), \phi =-\frac{2\pi}{3}
\end{eqnarray}
One can easily verify that this is also equivalent to a rotation along ${\bf n}=(0,1,0)$ axis and $\phi=\frac{\pi}{2}$ therefore the resultant {\em EFT} describes a $p-wave$ state with a $\pi/2$ phase shift.
The state therefore breaks the time reversal symmetry, physically distinct from a $T$-invariant state. The {\em EFT} for ${\tilde \chi}$ (here we rename them as $\chi$) is
\begin{widetext}
\begin{eqnarray}
H_{TS1} =\frac{1}{2} \int d{\bf r} [ \chi^T({\bf r}) (\tau_x \otimes (\sigma_x {i\nabla_z} -\sigma_z {i\nabla_x}) -\tau_z \otimes \mathbb{I} {i\nabla_y}) \chi ({\bf r}) 
+ g_1 \chi^T \tau_y \chi \chi^T  \tau_y  \chi +g_2 \chi^T  \tau_x\otimes \sigma_y \chi  \chi^T \tau_x\otimes \sigma_y  \chi ] 
\label{TSF1}
\end{eqnarray}
\end{widetext}
where $\chi^T= (\chi_{1\uparrow}, \chi_{1\downarrow}, \chi_{2\uparrow},\chi_{2\downarrow})$.

{\em B: Weyl Fermions} 

We then consider a rotation,

\begin{eqnarray}
{\bf n}=\frac{1}{\sqrt{3}}(1,1,1), \phi =\frac{2\pi}{3}.
\end{eqnarray}

This transformation leads a more surprising EFT, one that is related to a real fermion representation of interacting Weyl fermions but with {\em single} chirality; 
\begin{widetext}
\begin{eqnarray}
H_{Weyl} =\frac{1}{2} \int d{\bf r} [ \chi^T ({\bf r}) (\mathbb{I}  \otimes (\sigma_x  {i \nabla_x} +\sigma_z {i\nabla_z}) +\tau_y \otimes \sigma_y  {i\nabla_y}) \chi ({\bf r}) 
+ g_1 \chi^T \tau_x \otimes \sigma_y \chi \chi^T \tau_x\sigma_y \chi +g_2 \chi^T \tau_z \otimes \sigma_y \chi \chi^T \tau_z \otimes \sigma_y \chi ]. \nonumber \\
\label{Weyl}
\end{eqnarray}
\end{widetext}
This mapping was also applied to understand topological quantum criticality in previous studies\cite{Zhou22}.
Phenomenologically, the emergent symmetries we are going to discuss below can be easily related to a chiral symmetry in Weyl fermions but with a very distinct feature.
The standard $U(1)$ chiral anomalies in quantum field theories are absent and Weyl fermions as emergent particles implied in Eq.(\ref{Weyl})  have only one fixed chirality, either left or right but not both.
{\em Weyl fermions with one single chirality usually do not appear in three dimensional bulk Weyl metals so their appearance here in an EFT of gapless superfluids is a big surprise.}
Such single-Weyl-cone phenomena usually only occur on a 3D surface of a 4D lattice so are hypothetical from physics point of view. However, equivalent dynamics can be physical reality in gapless superfluids.

\subsection{$K$-rotations and mapping and $\tau-\sigma$ duality}

We now turn to a class of {\em EFTs} where $SU(2)$ rotation subgroup of $S^{ij}$ in Lorentz group $SO(3,1)$ is given by $F$-rotations instead of $K$-rotations in $Spin(4)$ group.
We will study the members generated by the dual group $S^{ij}_D$ which in this case is $SU_K(2)$ subgroup in $Spin(4)$. All $K$-rotations leave $S^{ij}$ and their rotation group $SU_F(2)$ or $F$-rotations invariant; all {\em EFT}s
here again are given by the same $S^{ij}$.

We can further apply $K$ rotations and generate mapping between {\em EFTs} for different physics systems. The discussions are very similar to those in the previous subsection and we simply list the results below.
They can also be related to $F$ rotations by a simple $\tau-\sigma$ duality transformation,

\begin{eqnarray}
\tau_i \rightarrow \sigma_i, \sigma_i \rightarrow \tau_i, i=x,y,z.
\end{eqnarray}

{\em C: Dual of Model A}

\begin{widetext}
\begin{eqnarray}
H_{DTS}=\frac{1}{2} \int d{\bf r} [ \chi^T ( (\tau_x  {i \nabla_x} -\tau_z {i\nabla_z})\otimes \sigma_z +1 \otimes \sigma_x  {i\nabla_y}) \chi
+ g_1 \chi^T \tau_y \otimes \sigma_z \chi \chi^T \tau_y \otimes \sigma_z \chi +g_2 \chi^T  1 \otimes \sigma_y \chi \chi^T 1 \otimes \sigma_y \chi ]; \nonumber \\
\label{DTS}
\end{eqnarray}
\end{widetext}

{\em C1: Dual of Model A1}

\begin{widetext}

\begin{eqnarray}
H_{DTS1}= \frac{1}{2} \int d{\bf r} [ \chi^T ( (\tau_x  {i \nabla_x} -\tau_z {i\nabla_z})\otimes \sigma_x -\sigma_z  \otimes 1 {i\nabla_y}) \chi
+ g_1 \chi^T \tau_y \otimes \sigma_x \chi \chi^T \tau_y \otimes \sigma_x \chi +g_2 \chi^T  1 \otimes \sigma_y \chi \chi^T 1 \otimes \sigma_y \chi ]; \nonumber \\
\label{DTS1}
\end{eqnarray}
\end{widetext}

{\em D: Dual Model of B as Gapless Nodal Phase with P-symmetry}

The $\tau-\sigma$ dual of Weyl fermions is exactly an EFT for gapless superconducting nodal phases with parity-symmetry.  It has the following explicit form,

\begin{widetext}
\begin{eqnarray}
H_{NP}=\frac{1}{2} \int d{\bf r} [ \chi^T  (\tau_x  {i \nabla_x} +\tau_z {i\nabla_z})\otimes \mathbb{I} +\tau_y \otimes \sigma_y  {i\nabla_y}) \chi
+ g_1 \chi^T \tau_y \otimes \sigma_x \chi \chi^T \tau_y\otimes \sigma_x \chi +g_2 \chi^T \tau_y \otimes \sigma_ z\chi \chi^T \tau_y \otimes \sigma_z \chi ]. \nonumber \\
\label{DW}
\end{eqnarray}
\end{widetext}

Indeed, $H_{NP}$ in Eq.(\ref{DW}) are related to $H_{DTS}$ and $H_{DTS1}$ via $F$-rotations, $U_F$ defined below

\begin{eqnarray}
U_F(&{\bf n} &, \theta) = \cos\frac{\phi}{2} -i\sin\frac{\phi}{2} (  n_x \tau_y \otimes \sigma_x +n_y \sigma_y +n_z \tau_y\otimes \sigma_z); \nonumber \\
&& {\bf n}=\frac{1}{\sqrt{3}}(1,1,1), \phi =\pm \frac{2\pi}{3}.
\end{eqnarray}

\end{document}